\documentclass{aastex61}

\received{}
\revised{}
\accepted{}

\submitjournal{ApJ}

\shorttitle{GA CME}
\shortauthors{Dai et al.}

\usepackage{bm}
\usepackage{threeparttable}
\usepackage{hyperref}
\usepackage[all]{hypcap}
\usepackage{amsmath}

\begin{document}

\title{Electron Density Reconstruction of Solar Coronal Mass Ejections Based on Genetic Algorithm: Method and Application}

\correspondingauthor{Xinghua Dai}
\email{xinghua.dai@outlook.com}

\author{Xinghua Dai}
\affiliation{School of Physics and Information Engineering, Shanxi Normal University, Linfen 041000, China}
\affiliation{Key Laboratory of Solar Activity, National Astronomical Observatories, Chinese Academy of Sciences, Beijing 100101, China}

\author{Huaning Wang}
\affiliation{Key Laboratory of Solar Activity, National Astronomical Observatories, Chinese Academy of Sciences, Beijing 100101, China}
\affiliation{University of Chinese Academy of Sciences, Beijing 100049, China}

\author{Bernd Inhester}
\affiliation{Max Planck Institute for Solar System Research, Justus-von-Liebig-Weg 3, 37077 G\"ottingen, Germany}


\begin{abstract}
We present a new method to reconstruct three dimensional electron density of Coronal Mass Ejection (CME) based on genetic algorithm, namely genetic reconstruction method (GRM). GRM is firstly applied to the model CMEs with different orientations and shapes. A set of analytic GL98 model CMEs are employed to produce synthetic CME images for GRM reconstruction. Model CMEs with longitude of $0^\circ$, $45^\circ$, $90^\circ$, $135^\circ$, $180^\circ$ and latitude of $0^\circ$, $15^\circ$, $30^\circ$, $45^\circ$ are used to test performance of GRM. The model CMEs are obscured with simulated occulter of coronagraph to find out the influence of incompleteness of CME brightness. We add random noise to some synthetic CME images to test the GRM performance. The CME reconstructions are carried out using synthetic data from STEREO A and B with separation angle of $90^\circ$ and from STEREO A and SOHO  with separation angle of $73^\circ$, respectively. Pearson correlation coefficient ($PCC$) and mean relative absolute deviation ($MRAD$) are calculated to analyse similarity between model and reconstructed CMEs for brightness and electron density. Comparisons based on the similarity analysis under various conditions stated above give us valuable insights of advantages and limitations of GRM reconstruction. Then the method is applied to real coronagraph data from STEREO-A, B and SOHO on September 30th, 2013.
\end{abstract}


\section{Introduction} 
\label{sec:intro}

Coronal mass ejections (CMEs) are usually observed by coronagraph \citep{1939MNRAS..99..580L}. The observation records radiation of Thomson scattering \citep{1930ZA......1..209M,1950BAN....11..135V,1966gtsc.book.....B,2006ApJ...642.1216V,2009SSRv..147...31H,2015arXiv151200651I} which is produced by interaction between radiation from the photosphere and free electrons inside CMEs. Information of the electron locations along line of sight (LOS) is hidden to the observer after the observation \citep{2008JGRA..113.1104H,2009SoPh..256..183T}. 

Those hidden information can be restored by different methods of CME reconstruction \citep{2010AnGeo..28..203M,2011JASTP..73.1156T}. We classify the various methods into three categories as follows. 

(1) Methods to reconstruct point  features of CMEs. The three dimensional (3D) coordinates of a CME feature can be calculated if the feature can be seen from different view  points like Solar Terrestrial Relations
Observatory (STEREO) A and B \citep{Kaiser2008,2008SSRv..136...67H}.  For example, \cite{2008SoPh..252..385M} calculated 3D position of the feature using height-time diagrams from COR1 onboard STEREO-A and B based on epipolar geometry \citep{2006astro.ph.12649I}. \cite{2010ApJ...710L..82L,2010ApJ...722.1762L,2014NatCo...5E3481L} obtained the 3D position of CME feature from corona to helioshpere using  time-elongation map from COR2, HI1 and HI2 onboard STEREO-A and B. 

(2) Methods to reconstruct outline of CMEs. A global outline can be obtained from the CME boundary measurements in the coronagraph images. For example, \cite{2004GeoRL..3121802P} using the CME boundaries to construct a series of quadrilaterals to approximate the three dimensional CME outline based on the synthetic STEREO images. \cite{2009SoPh..256..167D} applied this method to reconstruct the CME outline using coronagraph observations from STEREO. \cite{2006ApJ...652..763T,2009SoPh..256..111T} employed a forward model of Graduated Cylindrical Shell (GCS) to reconstruct flux rope-like CMEs using data from either single viewpoint or multiple viewpoints. \cite{2010NatCo...1E..74B} employed an elliptical tie-pointing technique to reconstruct a full CME front in 3D space. \cite{2012ApJ...751...18F,Feng2013} developed a 3D mask fitting reconstruction method using coronagraph images from three viewpoints to obtain the 3D morphology of a CME. 

(3) Methods to reconstruct density inside CMEs. For example, \cite{2004Sci...305...66M} derived 3D position of electrons along LOS corresponding to every CME pixel through polarization analysis of single-view images. \cite{2014ApJ...780..141D} suggested a classification of ambiguity in this method to improve the reliability of the reconstruction. \cite{2009SoPh..259..199A} reconstructed 3D CME electron density through combination of  inversion using PIXON method \citep{2005ARA&A..43..139P} and forward modeling \citep{2006ApJ...652..763T,2009SoPh..256..111T}. \cite{2009ApJ...695..636F} developed a  reconstruction method based on level-set \citep{2006JBO....11f4029J} algorithm using synthetic CME images produced from 2D slice of magnetohydrodynamic (MHD) simulation \citep{2008ApJ...684.1448M}. \cite{2017A&A...599A..68H} developed Automated CME Triangulation (ACT) to find the most probable location where the CME passes at height of 5 $R_\odot$. ACT is effective under different observations from three or two viewpoints over a sliding window of five hours.

Information of CME structure obtained from 3D reconstruction gradually increases by methods from category (1) to (3). For category (3), the reconstructed CME is more close to the real CME which is a collective of inhomogeneous magnetized plasmas. On the other hand, ambiguities of these methods in category (3) are also larger than those in category (1) and (2). In the current work, we develop a new method to reconstruct the 3D electron density of CME based on genetic algorithm (GA) \citep{1992Holland,2011JGRA..11611203H}, namely genetic reconstruction method (GRM). According to the classification mentioned above, GRM belongs to category (3). 

The rest of this paper is organized as follows. We introduce GRM in Section \ref{sec:reconstruction},  and present application of the method in Section \ref{Model_CME} and \ref{Observed_CME} for model and observed CMEs. Finally, conclusions are  given in Section \ref{sec:summary}.

\section{Genetic Reconstruction Method}
\label{sec:reconstruction}

\begin{figure}[ht!]
	\centering
	\includegraphics[width=6in]{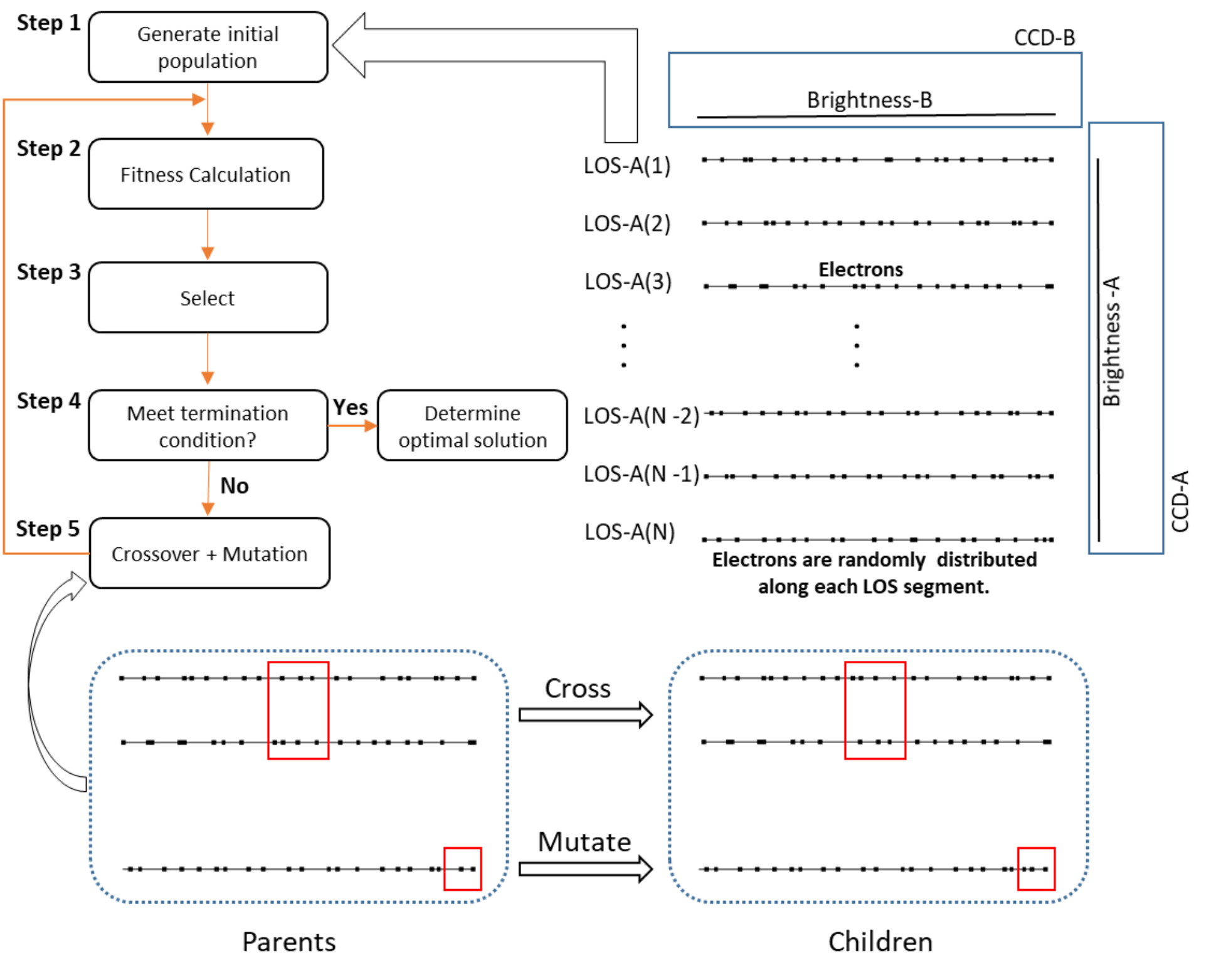}
	\caption{Flow chart of GA used in this work is shown in the top left panel. Individual of population is generated from the redistributions of the electrons along each LOS segment of STEREO-A as shown in the top right panel. The concepts of genetic operations, crossover and mutation, are shown in the bottom panel.\label{Fig_GA_flow}}
\end{figure}

Before method description in this section, we briefly explain the approach to reconstruct CME electron density using GRM. Our approach is to randomly distribute test electrons through a 3D reconstruction space, with their populations within voxels used to calculate line of sight integrations that create synthetic observations. These can be compared to the true observations, giving a goodness of fit to the electron distribution. We can randomly distribute electrons in the reconstruction space to create many different test models, and their goodness of fit to the data sets a scoring to the models. This sets a basis for a genetic algorithm approach to improve the fitness of synthetic brightness to observation. According to the scoring of fitness, we select the better electron distribution to the next iteration. When the iteration process meet the termination condition, we obtain the final reconstruction of electron distribution. Process of the reconstruction with five steps is shown in Figure \ref{Fig_GA_flow}.

\subsection{A general description for CME reconstruction}
\label{sec:General description}

Observed brightness, $B_{obs}$, of each CME pixel is the corresponding LOS integral of Thomson Scattering including measurement noise $\eta$:
\begin{equation}\label{eq_basic}
B_{obs} = \int_{\theta_{1}}^{\theta_{2}} B_e(\theta) N_e(\theta)d\theta + \eta.
\end{equation}
Here, $N_e(\theta)$ is the electron density on LOS where the angle between the plane of sky (POS) and the line connecting the electron to the Sun center has value $\theta$. The analytic function $B_e(\theta)$ \citep{2002A&A...393..295Q} is the value of Thomson scattering for a single electron at location $\theta$. $\theta_1$ and $\theta_2$ are LOS boundaries constrained by the CME shape.

CME reconstruction tries to obtain the vector \pmb{$N_e(\theta)$} when $B_{obs}$ is observable. Methods in category (1) focus on the localized feature who contributes most of the brightness along the LOS. Methods in category (2) focus on the electron density who contributes the brightness of the observed CME boundary. Methods in category (3) focus on the electron density along the LOS between $\theta_{1}$ and $\theta_{2}$. We summarize part of the challenges and limitations of CME reconstruction \citep{2009ApJ...695..636F,2010AnGeo..28..203M,2011JASTP..73.1156T} as follows:

(1) Equation \ref{eq_basic} is highly underdetermined. There are multiple possible solutions of Equation \ref{eq_basic} if constraints or a priori information are not enough. Possible distributions of the \pmb{$N_e(\theta)$} may generate the same observed brightness $B_{obs}$.

(2) Mix of CME and background corona. The CME images always contain brightness of background corona because of the optically thin nature of Equation \ref{eq_basic}, $B_{obs}$ = $B_{cme}$ + $B_{bg}$. $B_{bg}$ can be removed by subtraction of pre-CME image from CME image \citep{2010ApJ...722.1522V}. Situation becomes complex when the background corona contains dynamic structures like streamers, jets or other CMEs. It may be difficult to do a clean subtraction. Some advanced techniques are developed to make more effective background subtraction by more sophisticated manners \citep{2012ApJ...752..144M,2015ApJS..219...23M}.

(3) Measurement noise in the coronagraph images. The noise,  $\eta$, of any measurements is unavoidable. As pointed out by \cite{2011JASTP..73.1156T}, the presence of noise in the CME images may lead to unstable solutions of Equation \ref{eq_basic}. 

(4) Incomplete observation caused by occulter of coronagraph. The CME may be obscured by occulter of coronagraph  especially when the CME has not propagated far away from the Sun. And for different viewpoints, occulters may obscure different parts of the CME.

(5) The number of simultaneous observation viewpoints. Simultaneous observations from different viewpoints play important role to most of the reconstruction methods classified in Section \ref{sec:intro}. Inverse methods that try to restore \pmb{$N_e(\theta)$} usually need data from many view directions, like tomography. With two or three views, like Solar and Heliospheric Observatory (SOHO) \citep{Domingo1995,Brueckner1995} and STEREO, the inverse problem is extremely ill-posed. Solution of Equation \ref{eq_basic} is not unique.

(6) Separation angle between view directions. For reconstructions using data from two views, separation angle of $0^\circ$ and $180^\circ$ is unsatisfactory while $90^\circ$ is ideal. STEREO A and B orbit the Sun with different speed and the angle between them varies about $44^\circ$ per year.

We should keep these limitations in mind through out the process of CME reconstruction. 

CME reconstruction using inverse methods can be stated as an optimization problem:
\begin{equation}\label{eq_optimization}
cf = |\sum_{\theta_{1}}^{\theta_{2}}B_e(\theta)N_e(\theta) - B_{obs}|^p + regularization,
\end{equation}
where $cf$ is called cost function. $p$ is a positive number which defines the error norm of the observations. Traditionally, p=2 and the  regularization term also depends on $N_{e}^2$ so that minimizing Equation \ref{eq_optimization} results in a least-squared problem for which many solvers exist. According to the limitations of CME reconstruction that mentioned above, the optimization is highly ill-posed unless the regularization expression is chosen properly so that it yields different penalties for the many solutions in the null space of the data term.

As shown in Equation \ref{eq_optimization}, an appropriate regularization needs to be added to obtain more stable solution to deal with the ill-posed problem. Since many solutions are possible to eliminate the data term in Equation \ref{eq_optimization}, (in case that $\eta$=0, else the data $B_{obs}$ is probably inconsistent and there is no exact solution but there are many for which the delta error is of the order of $\eta$) the solution which is finally returned by any procedure depends on
the error norm and the regularization we choose. We emphasize that this ambiguity lies in the heart of the CME reconstruction problem and is unavoidable because two views are not sufficient to determine a unique CME
density distribution. Our goal is to test whether a certain combination of error norm p and regularization may yield more realistic solutions than others. Since conventional solvers only work for p=2 and $N_e^2$ dependent regularization, we decided to use GRM to find the minimum of Equation \ref{eq_optimization} so that we maintain a maximum of flexibility to choose p and the regularization expression.

In order to show the ability of GRM itself for CME reconstruction, the regularization has not been added in the current work and we choose p=1: 
\begin{equation}\label{eq_optimization2}
cf = |\sum_{\theta_{1}}^{\theta_{2}}B_e(\theta)N_e(\theta) - B_{obs}|.
\end{equation}
The L1 norm is often favoured over least-squared solutions in image processing because it attributes less weight to measurement outliers. A systematic test of different regularization operators will be made in the next paper.

\subsection{Calculations of electron number and white light brightness}
\label{sec:optimization}

The optimization is initiated by calculating electron number for each LOS and redistributing the electron density $N_e(\theta)$ in a discrete form along each LOS. Thomson scattering can then be calculated for the redistributed electrons and the cost function is ready for the optimization.

\begin{figure}[ht!]
	\centering
	\includegraphics[width=7in]{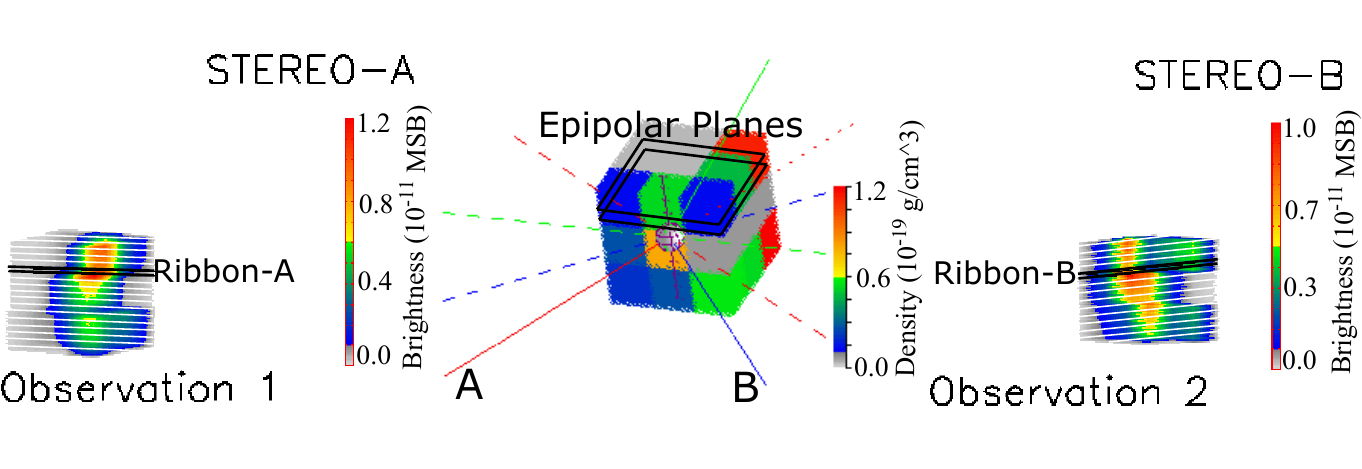}
	\caption{The density of CME plasmas is modeled within a cube as shown in the middle panel. The Sun is represented by a grey ball. Directions of STEREO A , B and the Earth are shown by red, blue and green solid lines, respectively. The corresponding planes of sky are shown by the dashed lines with the same colour. The simulated brightness in FOV of STEREO A and B are shown in the first and third panel. \label{Fig_GA_evolve_3D_0930_0008_art_new}}
\end{figure}

In order to explain the process of initiation, a cube of plasmas is used to mimic CME as shown in the middle panel of Figure \ref{Fig_GA_evolve_3D_0930_0008_art_new}.  In the first and third panel, the simulated white light brightness is produced through Thomson scattering mechanism in the field of view (FOV) of STEREO A and B, respectively. There is no need to subtract the background corona since we only simulate the brightness of model CME itself. For brightness of real CME in section \ref{Observed_CME}, we rescale the CME images to 128$\times$128 pixels and make a 3$\times$3 smoothing to minimize the influence of noise which is inherent in the measurements. A pre-CME background image is subtracted to get the excess brightness of CME \citep{2010ApJ...722.1522V}.

It is necessary to split the brightness for simplifying the reconstruction. As shown in Figure \ref{Fig_GA_evolve_3D_0930_0008_art_new}, the brightness is equally split up into ribbons with white lines for FOV A and B. For example, the corresponding part of Ribbon-A in FOV A is Ribbon-B in FOV B displayed by black lines. Pixels in this pair of ribbons belong to the same set of epipolar planes \citep{2006astro.ph.12649I} between two epipolar planes indicated in the middle panel of Figure \ref{Fig_GA_evolve_3D_0930_0008_art_new}. Process for finding ribbon B is carried out by projecting LOS of each pixel in ribbon A into FOV B. LOS A that is projected into FOV B should intersect with the CME boundary at least at two cross points unless the LOS is tangent to the CME boundary. Then the projection of LOS A and LOS A itself become finite line segments between those two cross points, not the infinite LOS from the observer to infinity. We assume that all the LOSs are parallel to each other in the FOV of a given telescope. Reconstruction can be carried out for each pair of the ribbons to obtain the electron density in the corresponding set of epipolar planes.

We use $i$ to index pixels along epipolar lines in FOV A, $j$ to index computation grid along LOS A, $k$ to index epipolar planes. In each epipolar plane, coordinate axes are along the view directions and usually non orthogonal. Pixels in FOV A can be indexed by $(i,k)$. Integration along LOS A is equivalent to summation of index $j$.

(1) First of all, electron number along LOS of corresponding pixel of the model CME brightness image in FOV A can be calculated as:
\begin{equation}\label{num_electrons}
N_{e}(i,k)=\frac{B_{obs}(i,k)}{B_{e}(\theta(i,k))}.
\end{equation}
where $B_{obs}(i,k)$ is brightness value of the pixel $(i,k)$. Thomson Scattering of a single electron, $B_{e}(\theta(i,k))$, can be calculated as
\begin{equation}\label{Ibe_ga}
B_{e}(\theta(i,k))=\frac{\pi \sigma}{2z^2}[2((1-u)C+uD)-((1-u)A+uB)\cos^2\theta(i,k)].
\end{equation}
$\theta$ is traditionally assumed to be zero \citep{2010ApJ...722.1522V} when the electron location along the LOS is unknown. z is the distance from the observer to the scattering location  along the LOS. Three dimensional reconstruction , like \cite{2004Sci...305...66M,2015ApJ...801...39D}, can obtain the $\theta$ value of each pixel to get a more precise calculation of the electron number as done in this paper. $\sigma$ is the Thomson scattering cross section of an electron and $u$ is the limb darkening coefficient. $A$, $B$, $C$ and $D$ are the van de Hulst coefficients~\citep{1950BAN....11..135V}.

In this work, electron number is always calculated from observation 1 as shown in Figure \ref{Fig_GA_evolve_3D_0930_0008_art_new}. In principle, observation 1 can be anyone of STEREO A, B or SOHO. Observation 2 is not restricted to STEREO-B but also possible for SOHO as shown in section \ref{Application}.  This enable us to carry out the reconstruction without data of STEREO B. 

(2) Redistribution of the electrons of each pixel along the corresponding LOS. As shown in the top right panel of Figure \ref{Fig_GA_flow}, series of LOS are generated from Brightness-A and the extent of LOS is restricted by the length of Brightness-B. Then we get the LOS segment. Each LOS segment is uniformly separated into $N_{j}$=64 computational grids. The voxels on the computational grids are initially populated with a random distribution of electrons. An array of random decimal, $L(i,j,k)$, are used to represent the electron distribution. 

We calculate electron number in the $j_{th}$ voxel on the grid point along the LOS segment by the following equation:
\begin{equation}\label{Ibe_ga}
N_{v}(i,j,k)=\frac{L(i,j,k)}{\sum_{j=1}^{N_{j}} L(i,j,k)}N_{e}(i,k).
\end{equation}

(3) Calculation of the Thomson scattering for the redistributed electrons. Equation \ref{Itot_ga} converts the electron density $N_{v}(i,j,k)$ into white light emission $B_{av}(i,j,k)$:
\begin{equation}\label{Itot_ga}
B_{av}(i,j,k)=B_{e}(\theta_{a}(i,j,k))N_{v}(i,j,k),
\end{equation}
where
\begin{equation}\label{Itot_Be}
B_{e}(\theta_{a}(i,j,k))=\frac{\pi \sigma}{2z^2}[2((1-u)c+uD)-((1-u)A-uB)\cos^2\theta_{a}(i,j,k)].
\end{equation}

Then we can do the summation for $j$ to get the total brightness of a CME pixel $(i,k)$ in FOV A:
\begin{equation}\label{BLOSA}
B_{a}(i,k)=\sum_{j=1}^{N_{j}}B_{av}(i,j,k).
\end{equation}
The total brightness of a CME pixel $(j,k)$ in FOV B can also be calculated by a similar summation for $i$:
\begin{equation}\label{BLOSB}
B_{b}(j,k)=\sum_{i=1}^{N_{i}}B_{bv}(i,j,k)
\end{equation}
using Equation \ref{Itot_ga} and \ref{Itot_Be} when values of $\theta_{a}$ and the van de Hulst coefficients in FOV A are replaced by values of $\theta_{b}$ and the van de Hulst coefficients in FOV B. We emphasize that each $LOS\_B$ contains electrons from different $LOS\_A$. The values of $\theta$ and the van de Hulst coefficients should be calculated for Equation \ref{Itot_ga} and \ref{Itot_Be} according to the electron locations in the coordinate system of FOV A and FOV B, respectively.

\subsection{CME reconstruction using genetic algorithm}
\label{sec:genetic algorithm}

Now, we can calculate the cost function for pixel $(i,k)$ in FOV A using Equation \ref{eq_optimization2}:
\begin{equation}\label{err_relat}
cf(i,k)=|B_{a}(i,k)-B_{obs}(i,k)|
\end{equation}
and the cost function for pixel $(j,k)$ in FOV B:
\begin{equation}\label{err_relat_b}
cf(j,k)=|B_{b}(j,k)-B_{obs}(j,k)|.
\end{equation}

Once the cost function is ready, the process of CME reconstruction can be described in the following five steps which are typical for GA as shown in the top left panel of Figure \ref{Fig_GA_flow}.

(1) Population Initiation. The redistributed electrons along all of the LOS segments constitute one of the individual of GA. The same process of random redistribution is carried out for 200 times to produce the initial population $L_{q}(i,j,k)$, $q = 1,2,3,...,N_{q}, N_{q}=200$. This is the beginning of GA as shown in the top left panel of Figure \ref{Fig_GA_flow}. 

\begin{figure}[ht!]
	\centering
	\includegraphics[width=3.5in]{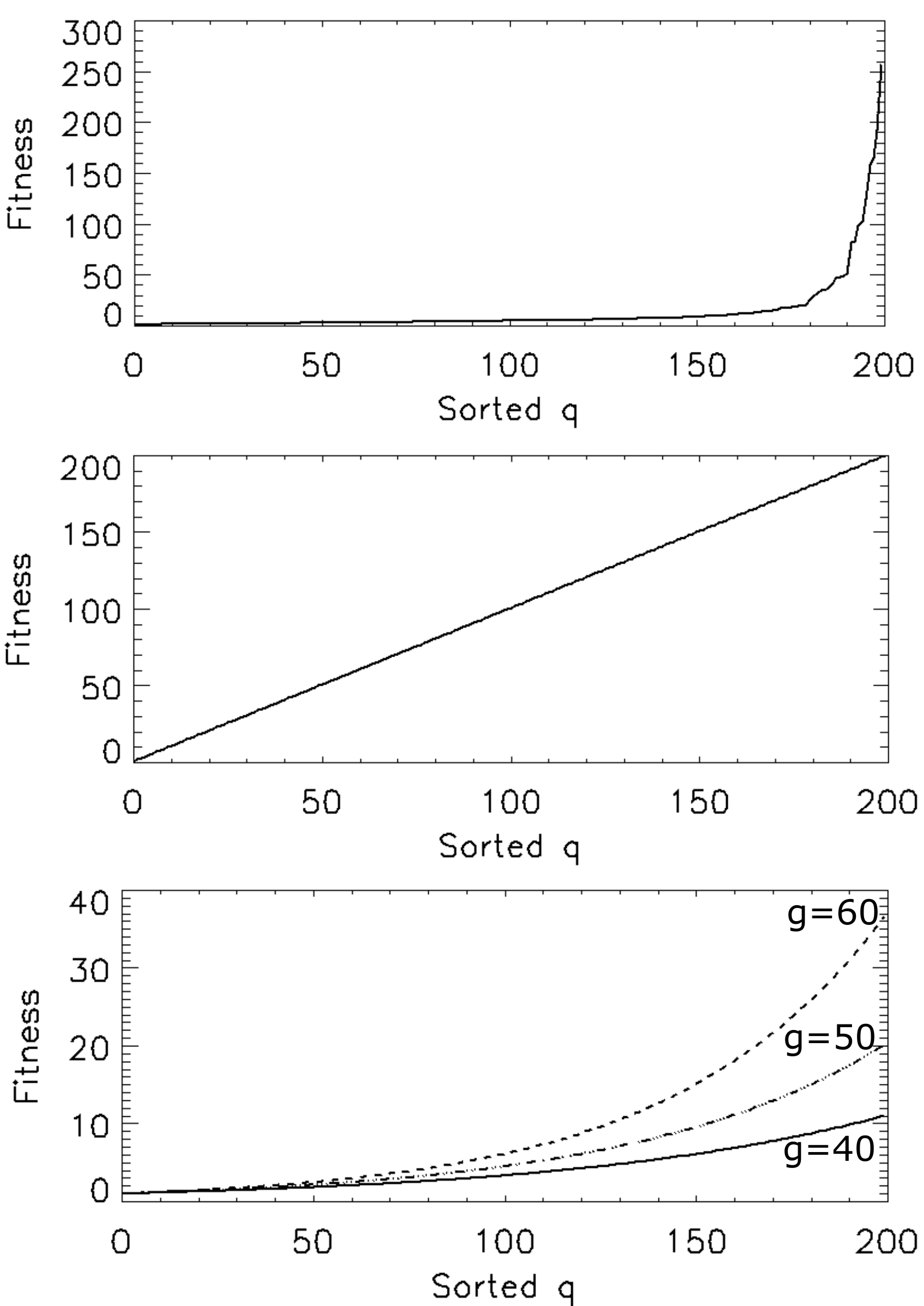}
	\caption{Top panel: Original fitness. Middle panel: Fitness transformed to arithmetic progression. Bottom panel: Fitness with natural exponential distribution. \label{Fig_fit_plot}}
\end{figure}

(2) Fitness calculation. Fitness function used in GA is inversely proportional to the cost function: 
\begin{equation}\label{fitness_0}
Fit_{q}=\frac{B_{obs}}{cf_{q}}.
\end{equation}
We show an example of the fitness of $N_{q}$ individuals in the top panel of Figure \ref{Fig_fit_plot}, from smallest to largest. In order to avoid local solution which may be produced by the steep distribution of fitness as shown in the top panel of Figure \ref{Fig_fit_plot}, we need to rescale the fitness function. Firstly, the values of fitness is changed to be linearly increasing by
\begin{equation}\label{fitness_sort}
Fit[sort(Fit)]=findgen(N_{q})+1,
\end{equation}
where $sort$ is the ranking function implemented in IDL. The sorted array of fitness $Fit[sort(Fit)]$ is incremental as shown in the top panel of Figure \ref{Fig_fit_plot}. The IDL function $findgen(N_{q})$ creates an arithmetic progression [0,1,2,...,$N_{q}-1$]. Then the values of fitness are transformed to [1,2,3,...,$N_{q}$] by Equation \ref{fitness_sort} as plotted in the middle panel of Figure \ref{Fig_fit_plot}. Finally, the IDL function $exp$ are used to transform the fitness values into natural exponential distribution by 
\begin{equation}\label{fitness_A}
Fit_{qg}(i,k)=exp(Fit_{q}(i,k)\frac{g}{N_{g}}0.03)
\end{equation}
and 
\begin{equation}\label{fitness_B}
Fit_{qg}(j,k)=exp(Fit_{q}(j,k)\frac{g}{N_{g}}0.03), 
\end{equation}
where $g=1,2,3,...,N_{g}, N_{g}=100$ is the total number of generation in the genetic evolution. As shown in the bottom panel of Figure \ref{Fig_fit_plot}, the fitness distribution becomes steeper from generation 40 to 60. It means that the algorithm can avoid local solution at the preliminary stage of genetic evolution and keep it more convergent at the later stage.

\begin{figure}[ht!]
	\centering
	\includegraphics[width=4.5in]{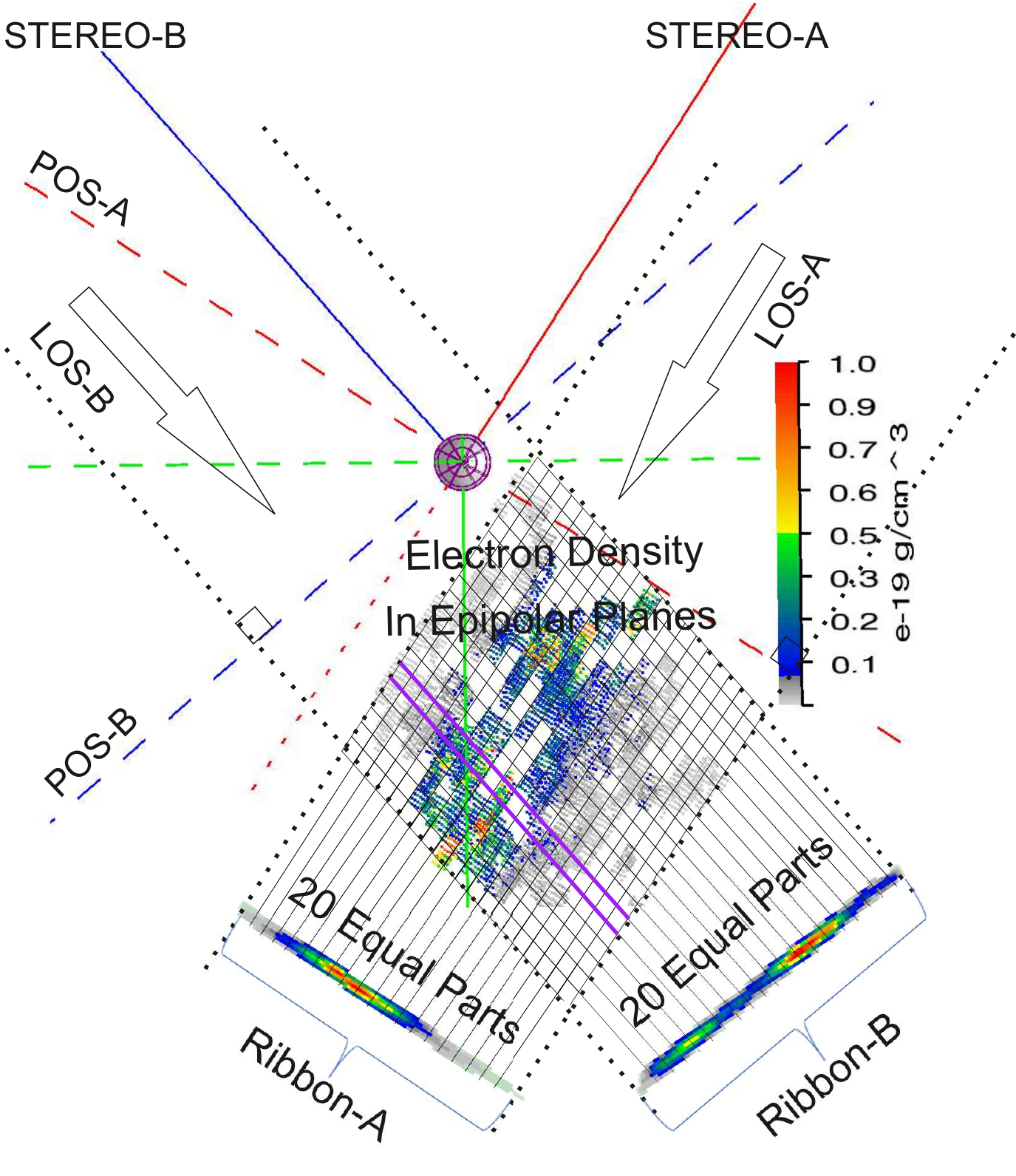}
	\caption{Genetic reconstruction of a pair of  brightness ribbons in FOV A and B. The reconstructed brightness ribbon A and B are the LOS integration of emission from the same set of epipolar planes as indicated in the middle panel of Figure \ref{Fig_GA_evolve_3D_0930_0008_art_new}. Ribbon A and B are equally split up into 20 parts respectively. Parallelograms are used to denote the 20$\times$20 equal parts of electron distribution. For each part of the brightness ribbon, there are 20 parts of electron distribution with shape of parallelogram to produce the emission along the corresponding LOS inside the purple column for instance. \label{Fig_GA_evolve_northmass_grid_0930_0008_art}}
\end{figure}

(3) Selection operator is employed to obtain the optimal electron distribution. In order to further simplify the reconstruction, we equally divide the brightness ribbon into 20 parts for FOV A and B respectively as shown in Figure \ref{Fig_GA_evolve_northmass_grid_0930_0008_art}. For each part of the brightness ribbon in FOV B, a group of electron distribution $L_{q}$ along the LOS segments inside a column which is marked by the purple lines for instance, contribute to the brightness of Thomson scattering. Mean value of fitness is already calculated for pixels inside each part of the brightness ribbon using Equation \ref{fitness_B} in step (2). We randomly choose two candidates from $N_q$ candidates of $L_{q}$ group and compare their mean fitness. The candidate with larger fitness will be chosen and passed on to the next generation. Such tournament is repeated $N_q$ times for each part of brightness ribbon in FOV B to produce a new population. based on this new population, brightness in FOV A can be updated and the mean fitness of the whole brightness ribbon in FOV A can be calculated by Equation \ref{fitness_A} for each of the $N_q$ individuals. Such mean fitness of the whole brightness ribbon is also calculated for FOV B based on the new population. Ten individuals with worst fitness in FOV B are replaced by ten individuals with best fitness in FOV A. This replacement optimizes the population for FOV A after the tournament selection for FOV B.

(4) In this work, we set the maximum number of generation to 100. If the genetic evolution meets this termination condition, the GA will stop and an optimal solution will be chosen to be the final reconstruction of CME. If the number of generation is less than 100, the algorithm will go to step (5).

(5) Crossover and mutation operators are used to update the population $L_{q}$. The basic concept of these operations is presented in the bottom panel of Figure \ref{Fig_GA_flow}. The crossover operator exchanges the electron distribution between two randomly selected individuals while the mutation operator changes the distribution inside one individual. Formula \ref{e_cross} and \ref{e_mutate} describe these operators:
\begin{equation}\label{e_cross}
L_{m}(i,j,k) \rightleftarrows  L_{n}(i,j,k),  
\end{equation}
where $L_{m}(i,j,k) \in CME_{m}$, $L_{n}(i,j,k) \in CME_{n}$ and $m,n \in [1,N_q], m\not= n$.
\begin{equation}\label{e_mutate}
L_{m}(i,j,k) =  L_{m}(i,j,k) + L_{m}(i,j,k) \times D_{r},  
\end{equation}
where $D_{r}$ is a random decimal which can be positive or negative. In order to simplify the reconstruction, crossover and mutation operators are applied to all of the $L_q$ inside one of the 20$\times$20 areas with shape of parallelogram at each time.

After the genetic operation, we get the new CME population from parents to children. Probabilities of crossover and mutation decrease progressively to keep the optimization being global and convergence.

Process from (2) to (5) are repeated until the number of generation is equal to $N_g$.

In the following sections, GRM is firstly applied to a set of model CMEs with different directions and shapes. Then the method is employed to reconstruct a CME observed by  coronagraphs of SOHO and STEREO.

\section{Application}
\label{Application}
In order to test the GRM method in a more realistic way, we use a set of analytic GL98 \citep{1998ApJ...493..460G} model CMEs with various orientations and shapes, as well as occultation to produce synthetic coronagraph data instead of the cube of plasmas as used in Figure \ref{Fig_GA_evolve_3D_0930_0008_art_new} at section \ref{sec:optimization}. The GL98 model constructs a flux rope to create typical three-part CME structure. The flux rope is an analytical solution of magnetohydrostatic (MHS) equation
\begin{equation}\label{mhs_eq}
\frac{1}{4\pi}(\nabla\times\bm{B})\times\bm{B} - \nabla p - \rho\bm{g} =0
\end{equation}
and solenoidal condition of magnetic field, $\nabla\cdot\bm{B} = 0$. The explicit solution of Equation \ref{mhs_eq} can be derived from solution of 
\begin{equation}\label{mhs_eq2}
\frac{1}{4\pi}(\nabla\times\bm{B})\times\bm{B} - \nabla p =0
\end{equation}
by applying a mathematical stretching transformation $r \rightarrow r - a$ to an axisymmetric sphere of magnetic flux \citep{1956PNAS...42....1C, 1956PNAS...42....5C} with radius $r_0$. The sphere center is located at $r_{1}$ from solar center. After the transformation, the magnetic flux appears to be a tear drop. Then the magnetic field of GL98 model can be expressed by a Bessel function and a free parameter, $a_{1}$ \citep{1956ApJ...123..498P, 1995ApJ...446..877L}. As pointed out by \cite{1998ApJ...493..460G}, the solution of Equation \ref{mhs_eq} can even be used to solve MHD equation 
\begin{equation}\label{mhd_eq}
\frac{1}{4\pi}(\nabla\times\bm{B})\times\bm{B} - \nabla p - \rho\bm{g} = \rho\left[\frac{\partial \bm{v}}{\partial t} + (\bm{v}\cdot\nabla)\bm{v}\right]
\end{equation}
under self-similar theory \citep{1984ApJ...281..392L}. GL98 model is implemented in Space Weather Modeling Framework (SWMF) \citep{2005JGRA..11012226T,2012JCoPh.231..870T} and is successfully used in studies of CME simulation \citep{2004JGRA..109.1102M, 2004JGRA..109.2107M, 2014JGRA..119.5449M, 2014PPCF...56f4006M,2005ApJ...627.1019L,2016ApJ...820...16J,2017ApJ...834..172J}. Recently, this model is applied in a user-friendly tool named Eruptive Event Generator using Gibson-Low configuration (EEGGL) \citep{2017ApJ...834..173J,2017JGRA..122.7979B} which has been transitioned to the Community Coordinated Modeling Center (CCMC).

Finally, we obtain the distribution of plasmas density frozen into magnetic field:
\begin{equation}\label{rho_eq}
\rho = f(B_{r},B_{\theta},B_{\phi}).
\end{equation}
We change position of the sphere center, $r_{0}$, $r_{1}$, $a$ and $a_{1}$ to obtain model CMEs with different orientation, size, shape and strength of magnetic field. Since the distribution of plasmas density is known, the synthetic CME white light images can be produced through the Thomson scattering mechanism mentioned in section \ref{sec:optimization}.

We apply GRM to these synthetic white light images to reconstruct the model CMEs. Comparisons of brightness and electron density between models and reconstructed CMEs are presented to show advantages and limitations of GRM in section \ref{Model_CME}. After systematic comparisons between model and reconstructed CME, GRM is applied to real observations from SOHO and STEREO in section \ref{Observed_CME}.

\subsection{Application for model CMEs}
\label{Model_CME}
\begin{figure}[ht!]
	\centering
	\includegraphics[width=4in]{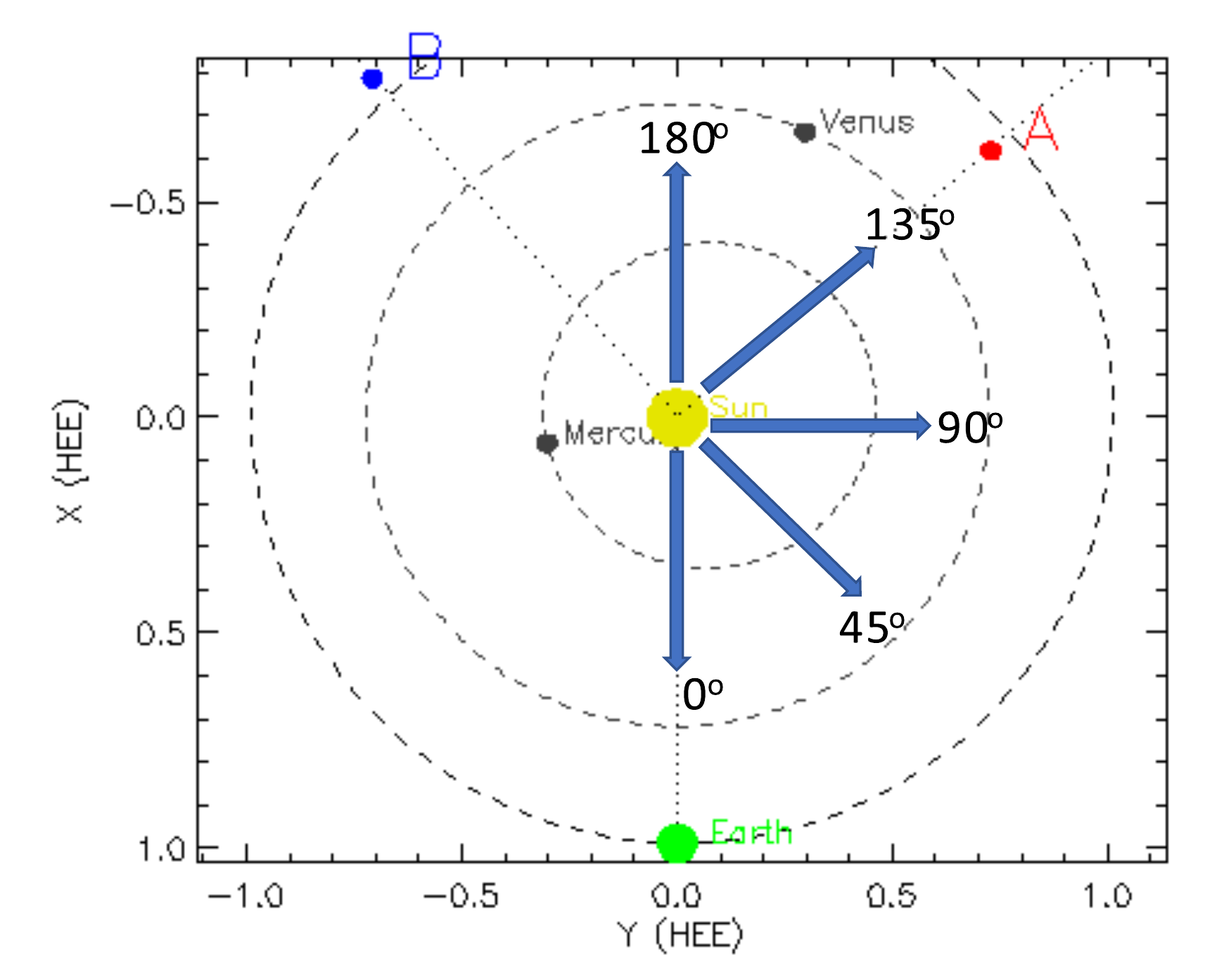}
	\caption{Positions of the Earth, STEREO-A and B on 2013 Feb 15 are shown with green, red and blue points. Longitudes of the GL98 model CMEs are illustrated by blue arrows marked with $0^\circ$, $45^\circ$, $90^\circ$, $135^\circ$ and $180^\circ$.\label{Fig_ABS_position_90}}
\end{figure}

We simulate CME observations using the same parameters of STEREO/COR2 and SOHO/C3 including satellite position, view direction, size of FOV and spatial resolution on February 15th, 2013. Positions of the Earth (SOHO), STEREO A and B in our simulation are shown in Figure \ref{Fig_ABS_position_90}.

For simulations of two view observations from STEREO A and B as observation 1 and 2, the separation angle is set to $90^\circ$ to test interfering factors of reconstruction like longitude and latitude of CME, whether or not to use occultation, whether or not the CME is halo. For STEREO A and SOHO as observation 1 and 2, the separation angle is set to $135^\circ$. We add data of COR2 from STEREO B as the third view to improve the performance of GRM.

Central longitude of the model CME is set to $0^\circ$, $45^\circ$, $90^\circ$, $135^\circ$ and $180^\circ$ respectively as shown in Figure \ref{Fig_ABS_position_90}. Central latitude of the model CME is routinely set to $0^\circ$ and $30^\circ$ for each value of longitude. Latitude of $15^\circ$ and $45^\circ$ are also set to the model CME with longitude of $0^\circ$ to test latitude dependence of GRM more carefully. For convenience, we mark the CME with (longitude,latitude) for specific values of longitude and latitude. For example, CME with longitude of $0^\circ$ and latitude of $0^\circ$ can be called (0,0) CME.

In order to quantitatively evaluate the similarity between the model and reconstructed CME, we calculate mean relative absolute deviation ($MRAD$) 
\begin{equation}\label{MRAD_eq}
MRAD=\frac{1}{N}\sum_{i=1}^{N}\frac{|x_i-y_i|}{x_i}
\end{equation}
and linear Pearson
correlation coefficient ($PCC$) \citep{2018JGRA..123.7257L}
\begin{equation}\label{PCC_eq}
PCC=\frac{\sum_{i=1}^{N}(x_i-\overline{x})(y_i-\overline{y})}{\sqrt{\sum_{i=1}^{N}(x_i-\overline{x})^2}\sqrt{\sum_{i=1}^{N}(y_i-\overline{y})^2}},
\end{equation}
where $x$ and $y$ are values of brightness or electron density of the model and reconstructed CME, $N$ is the number of CME voxel. If y is similar to x, value of $MRAD$ should be close to 0 and value of $PCC$ should be close to 1. The upper limit for $MRAD$ is set to 1.0.

We show similarity between the model and reconstructed CMEs qualitatively with images and quantitatively with $PCC$ in Figure \ref{Fig_0_0} and the following figures. For convenience, we name these figures ``Figure of Similarity" (FOS). We run GRM three times for each model CME to prove consistency of the method as shown in FOS which is labelled with ``1st", ``2nd" and ``3rd". Because of the ill-posed property of CME reconstruction and inherent randomness of GRM, reconstruction results including Thomson scattering brightness and electron density of the same model CME should be slightly different between each run.

For each run of GRM, we not only show the 3D distribution of electron density in the first column of last three rows of FOS but also the 2D distribution of density averaged along directions of x, y and z axis as shown in the last three columns of these rows. For example, the density averaged along direction of z forms the 2D distribution in xy plane as shown in the second column which is labelled with ``xy" in the lower left corner. Performance of GRM is different for a specific CME model on xy, xz and yz plane. Average density in these planes enable us to make more comprehensive comparisons between the model and reconstructed CME.

For comparison of brightness, we display the original brightness of GL98 model CME in the first column of the first and second row of FOS. The brightness reconstructed by GRM at generation 100 is labelled with ``1st", ``2nd" and ``3rd" for different runs of GRM. We plot $PCC$ of brightness from generation 0 to 100 in the third row to show the process of genetic evolution. The plots of $PCC$ at ``1st", ``2nd" and ``3rd" runs for reconstruction 1 are shown in the first column with colour of red, green and blue. Plots of $PCC$ for reconstruction 2 are shown in the second column. $PCC$ becomes larger from generation 0 to 100, which means that the optimization is convergent.

\begin{figure}[ht!]
	\centering
	\includegraphics[width=5.5in]{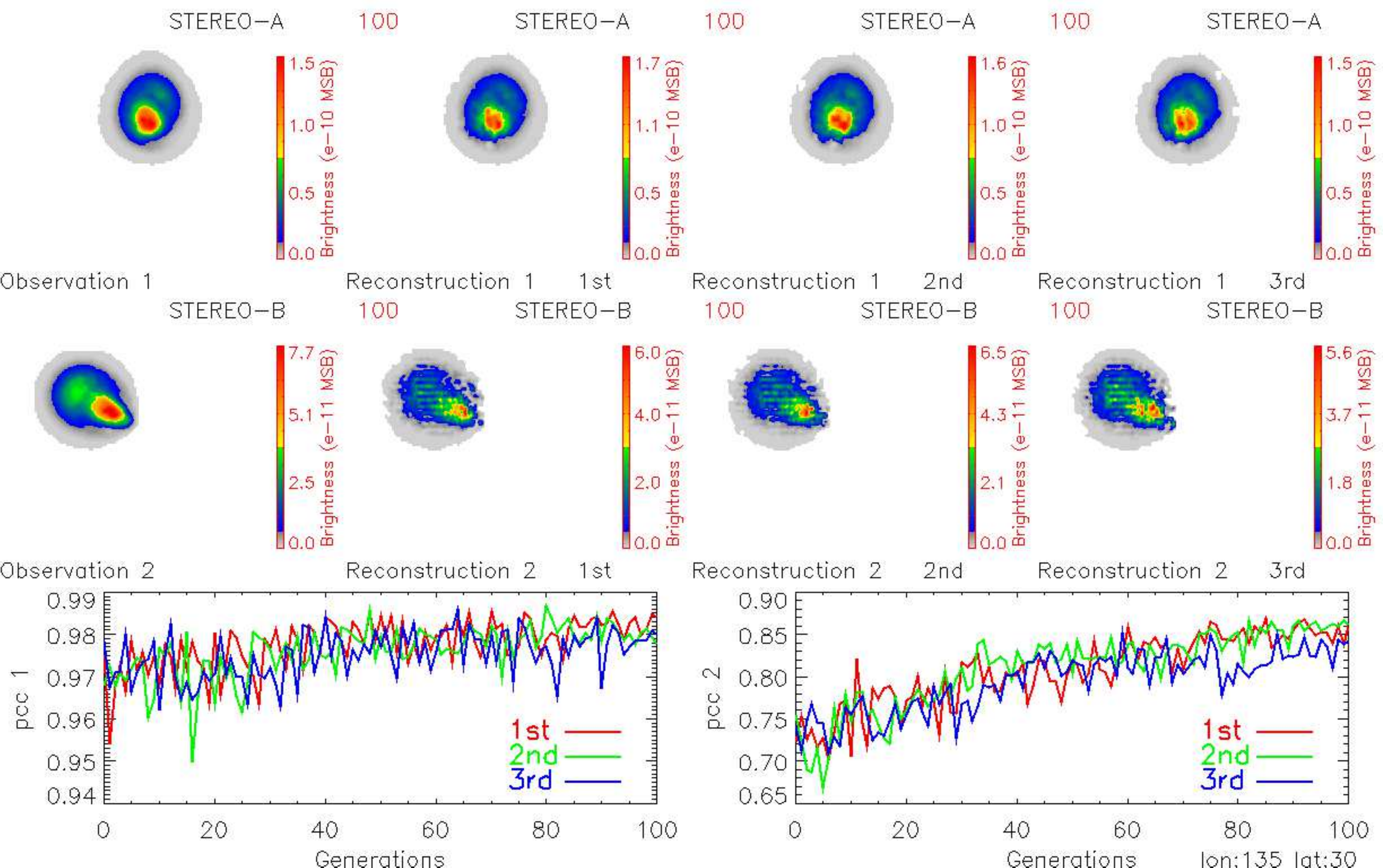}
	\\*
	\includegraphics[width=5.5in]{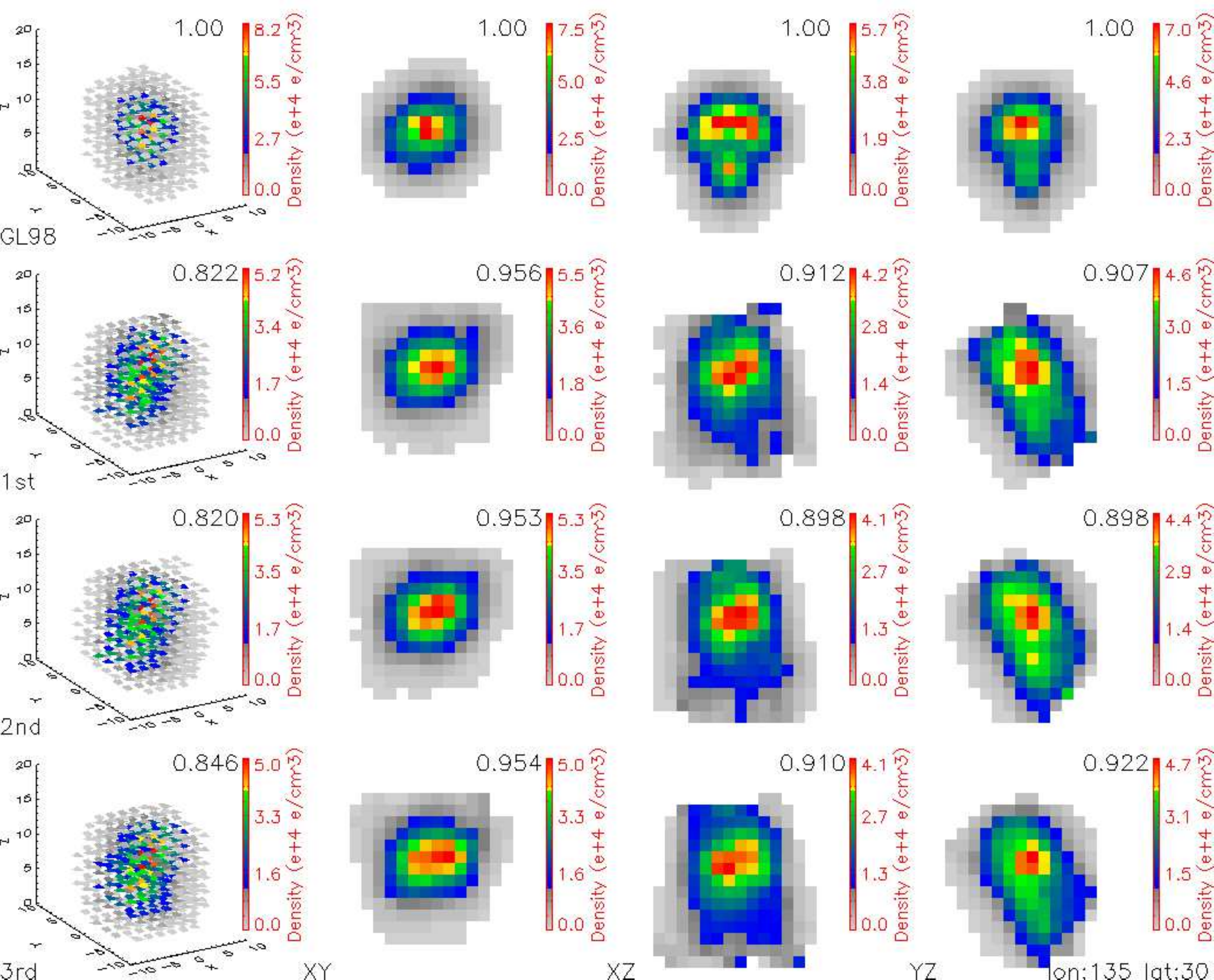}
	\caption{FOS of (135,30) CME. \hyperref[table_PCC]{Return to Table \ref{table_PCC}.}}\label{Fig_135_30}
\end{figure}

For comparison of electron density,  the electron density in 3D space as shown in FOS is the average inside cubes with width of 2.0 $R_{sun}$. The original density of GL98 model CME in 3D space, xy, xz and yz plane are firstly presented in the fourth row of FOS. Then reconstructed density of the ``1st", ``2nd" and ``3rd" runs of GRM are displayed in the last three rows of FOS. Heliocentric Earth Equatorial (HEEQ) \citep{2006A&A...449..791T} coordinate system is used to illustrate the CME density in FOS. The corresponding heliographic coordinates is Stonyhurst. In Figure \ref{Fig_0_0}, longitude and latitude of the model CME are both equal to $0^\circ$. It means that direction of the (0,0) CME is along z axis. In order to simulate the brightness of (135,30) CME, longitude and latitude of the model CME are changed to $135^\circ$ and $30^\circ$ respectively. Based on the simulated brightness, we reconstruct the electron density of CME using GRM. Then we transform the model and reconstructed CME by $-135^\circ$ and $-30^\circ$ for longitude and latitude respectively. After this kind of transformation, the density of model and reconstructed CME in Figure \ref{Fig_135_30} can be viewed under the same perspective as in Figure \ref{Fig_0_0}. Similar transformations are applied to the rest of FOS. $PCC$ of electron density distribution between model and reconstructed CME labelled in FOS are summarized in Table \ref{table_PCC}. In order to give a more complete picture of the GRM performance, values of $MRAD$ along with $PCC$ are shown in the left columns of Table \ref{table_PCC}.

\begin{deluxetable*}{ccCrlc}[ht!]
	\label{table_PCC}
	\tablecaption{$MRAD$ and $PCC$ of electron density between model and reconstructed CMEs. Please click on the blue values of central longitude and latitude at the first column to go to corresponding FOS.}
	\tablecolumns{6}
	\tablewidth{0pt}
	\tablehead{
		\colhead{Modeled CME position} &
		\colhead{$MRAD$ of xyz space} &
		\colhead{$PCC$ of xyz space} & 
		\colhead{$PCC$ of xy plane} & 
		\colhead{$PCC$ of xz plane} &
		\colhead{$PCC$ of yz plane} \\
		\colhead{(longitude, latitude)} &
		\colhead{1st  ,  2nd  ,  3rd} &
		\colhead{1st  ,  2nd  ,  3rd} & 
		\colhead{1st  ,  2nd  ,  3rd} & 
		\colhead{1st  ,  2nd  ,  3rd} &
		\colhead{1st  ,  2nd  ,  3rd}
	}
	\startdata
	\hyperref[Fig_0_0]{0, 0} & 0.612, 0.620, 0.592 & 0.784, 0.782, 0.780 & 0.848, 0.883, 0.899 & 0.634, 0.642, 0.666 & 0.957, 0.960, 0.949 \\ 
	0, 15 & 0.615, 0.608, 0.622 & 0.799, 0.798, 0.816 & 0.905, 0.891, 0.886 & 0.750, 0.751, 0.779 & 0.933, 0.949, 0.935 \\ 
	\hyperref[Fig_0_30]{0, 30} & 0.652, 0.651, 0.666 & 0.808, 0.795, 0.800 & 0.892, 0.883, 0.864 & 0.853, 0.851, 0.851 & 0.965, 0.964, 0.959 \\ 
	0, 45 & 0.744, 0.741, 0.732 & 0.796, 0.800, 0.793 & 0.841, 0.849, 0.817 & 0.864, 0.871, 0.863 & 0.962, 0.964, 0.965 \\ \hline
	45, 0 & 0.569, 0.577, 0.569 & 0.845, 0.870, 0.862 & 0.923, 0.915, 0.925 & 0.762, 0.805, 0.796 & 0.992, 0.991, 0.991 \\ 
	\hyperref[Fig_45_0_cut]{45, 0}\tablenotemark{a} & 0.640, 0.635, 0.632 & 0.737, 0.772, 0.759 & 0.703, 0.786, 0.739 & 0.502, 0.586, 0.562 & 0.988, 0.988, 0.986 \\ 
	45, 30 & 0.631, 0.630, 0.620 & 0.840, 0.823, 0.809 & 0.857, 0.850, 0.831 & 0.846, 0.803, 0.798 & 0.991, 0.990, 0.989 \\ \hline
	90, 0 & 0.608, 0.599, 0.612 & 0.810, 0.789, 0.799 & 0.938, 0.944, 0.924 & 0.665, 0.580, 0.652 & 0.946, 0.937, 0.927 \\ 
	90, 30 & 0.651, 0.668, 0.656 & 0.782, 0.798, 0.787 & 0.893, 0.890, 0.908 & 0.843, 0.802, 0.860 & 0.931, 0.926, 0.946 \\ \hline
	135, 0 & 0.592, 0.597, 0.581 & 0.838, 0.868, 0.855 & 0.980, 0.975, 0.977 & 0.685, 0.679, 0.726 & 0.923, 0.921, 0.900 \\ 
	\hyperref[Fig_135_0_cut]{135, 0}\tablenotemark{a} & 0.578, 0.589, 0.579 & 0.821, 0.811, 0.817 & 0.790, 0.779, 0.782 & 0.649, 0.654, 0.653 & 0.927, 0.884, 0.932 \\ 
	\hyperref[Fig_135_30]{135, 30} & 0.621, 0.618, 0.611 & 0.822, 0.821, 0.847 & 0.957, 0.953, 0.955 & 0.913, 0.899, 0.910 & 0.908, 0.898, 0.923 \\ 
	\hyperref[Fig_135_30_cut]{135, 30}\tablenotemark{a} & 0.630, 0.643, 0.620 & 0.813, 0.835, 0.840 & 0.945, 0.953, 0.955 & 0.888, 0.872, 0.899 & 0.874, 0.886, 0.896 \\
	\hyperref[Fig_135_30_noise_1]{135, 30}\tablenotemark{b} & 0.740, 0.727, 0.711 & 0.795, 0.801, 0.814 & 0.820, 0.881, 0.862 & 0.843, 0.838, 0.873 & 0.851, 0.826, 0.854 \\ 
	135, 30\tablenotemark{c} & 0.894, 0.897, 0.898 & 0.206, 0.398, 0.408 & 0.507, 0.284, 0.414 & 0.252, 0.499, 0.433 & 0.162, 0.381, 0.438 \\ 
	135, 30\tablenotemark{d} & 0.999, 0.999, 0.999 & 0.139, 0.184, 0.164 & 0.231, 0.139, 0.132 & 0.157, 0.218, 0.230 & 0.204, 0.112, 0.170 \\ 
	\hyperref[Fig_135_30_al]{135, 30}\tablenotemark{e} & 0.719, 0.710, 0.705 & 0.688, 0.731, 0.709 & 0.851, 0.857, 0.868 & 0.762, 0.773, 0.740 & 0.759, 0.763 0.744 \\ 
	\hyperref[Fig_135_30_alb]{135, 30}\tablenotemark{f} & 0.750, 0.761, 0.755 & 0.728, 0.716, 0.704 & 0.871, 0.869, 0.868 & 0.786, 0.822, 0.833 & 0.773, 0.773, 0.781 \\  \hline
	180, 0 & 0.596, 0.610, 0.608 & 0.789, 0.736, 0.788 & 0.871, 0.790, 0.891 & 0.653, 0.562, 0.649 & 0.958, 0.949, 0.939 \\ 
	180, 30 & 0.628, 0.655, 0.643 & 0.773, 0.792, 0.821 & 0.880, 0.856, 0.884 & 0.759, 0.775, 0.812 & 0.965, 0.953, 0.956 \\  \hline
	\hyperref[Fig_abl_real]{ABS}\tablenotemark{g} & 0.000, 0.235, 0.252 & 1.000, 0.874, 0.836 & 1.000, 0.913, 0.927 & 1.000, 0.835, 0.791 & 1.000, 0.892, 0.780 \\ 
	\hyperref[Fig_alb_real]{ASB}\tablenotemark{h} & 0.000, 0.323, 0.299 & 1.000, 0.801, 0.837 & 1.000, 0.854, 0.936 & 1.000, 0.797, 0.847 & 1.000, 0.803, 0.898 \\ 
	\enddata
	\tablenotetext{a}{The CME is obscured by the modelled occulter of coronagraph.} 
	\tablenotetext{b}{Randomly-distributed noise are added to the synthetic CME observations.}
	\tablenotetext{c}{Randomly-distributed noise $\times$10  are added to the synthetic CME observations.}
	\tablenotetext{d}{Randomly-distributed noise $\times$100 are added to the synthetic CME observations.}
	\tablenotetext{e}{The CME is reconstructed from modelled coronagraph images of STEREO A and SOHO as observation 1 and 2.} 
	\tablenotetext{f}{The CME is reconstructed from modelled coronagraph images of STEREO A, SOHO and B as observation 1, 2 and 3.} 
	\tablenotetext{g}{The CME is reconstructed from real coronagraph images of STEREO A, B and SOHO as observation 1, 2 and 3.} 
	\tablenotetext{h}{The CME is reconstructed from real coronagraph images of STEREO A, SOHO and B as observation 1, 2 and 3.} 
\end{deluxetable*}

We further discuss the interfering factors of GRM based on FOS and Table \ref{table_PCC} as follows.

\begin{figure}[ht!]
	\centering
	\includegraphics[width=5.5in]{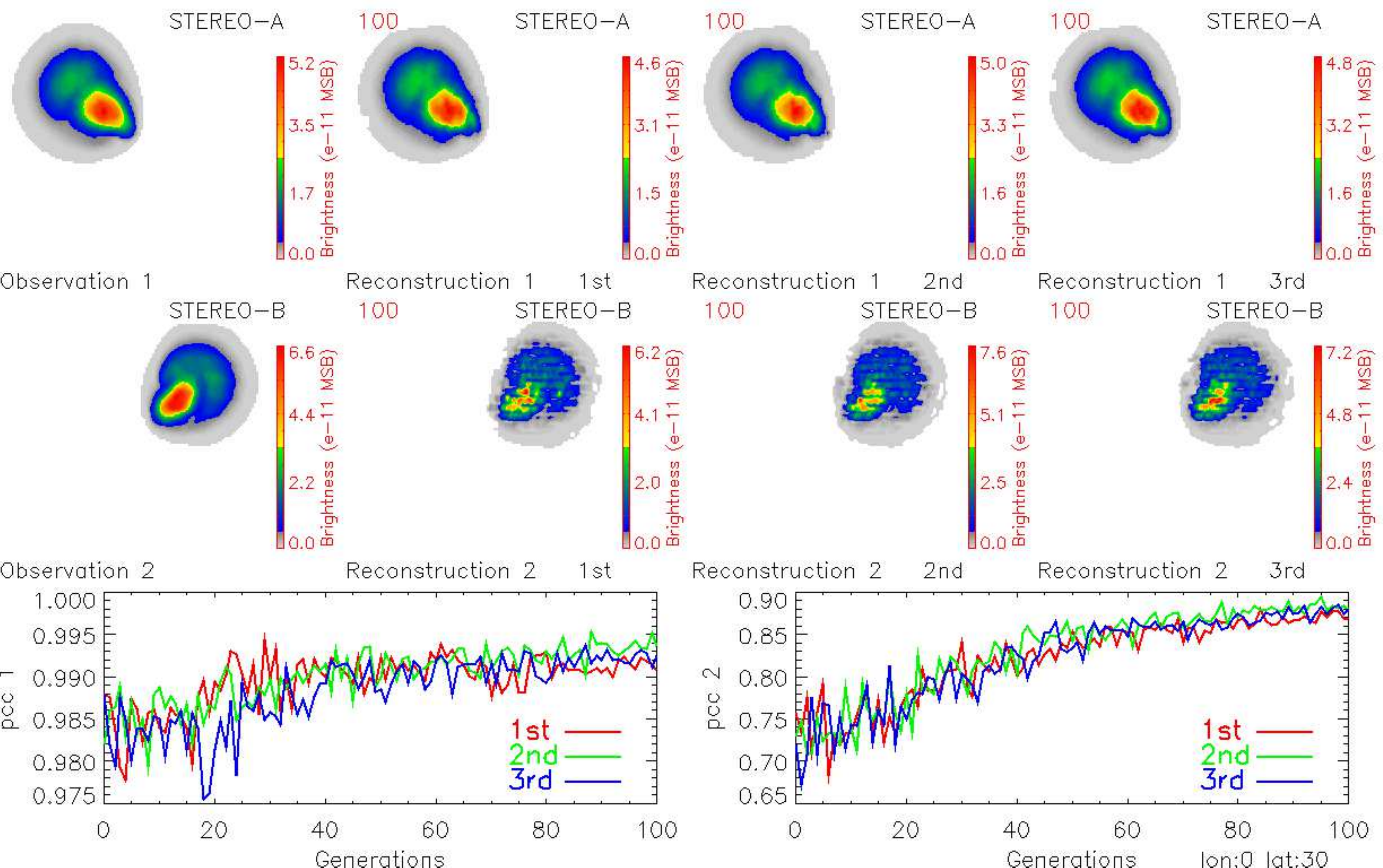}
	\\*
	\includegraphics[width=5.5in]{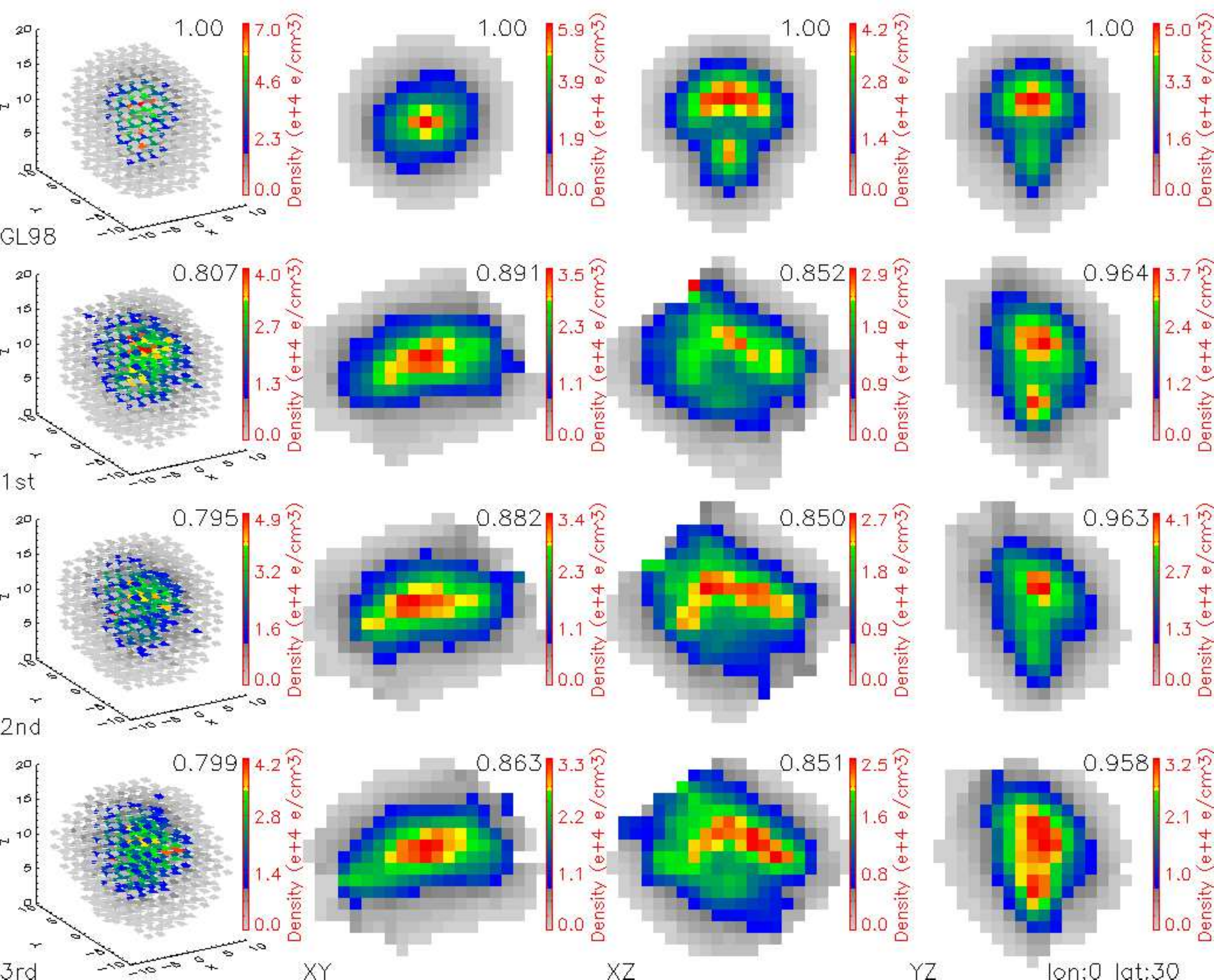}
	\caption{FOS of (0,30) CME. \hyperref[table_PCC]{Return to Table \ref{table_PCC}.}}\label{Fig_0_30}
\end{figure}

(1) As shown in Table \ref{table_PCC} and FOS, values of $PCC$ in xz plane is usually lower then those in xy and yz plane when latitude is equal to $0^\circ$. Value of $PCC$ in xz plane is even less than 0.6 at the second run for (90,0) CME. An obviously improvement of $PCC$ in xz plane can be seen when latitude grows from $0^\circ$ to $30^\circ$ for CMEs with all values of longitude. Comparison between (0,0) CME in Figure \ref{Fig_0_0} and (0,30) CME in Figure \ref{Fig_0_30} validate such improvement as an example. For latitude from $30^\circ$ to $45^\circ$, the improvement of $PCC$ stops as shown for the $0^\circ$ longitude CME.

$PCC$ of the (135,30) CME are the best among all of the reconstructed CMEs. For this CME, $PCC$ in 3D space are larger than 0.8 while $PCC$ in the xy, xz and yz plane are even larger than 0.9.

(2) The (45,0) CME appears full halo in FOV B as a back side event and the (135,0) CME appears full halo in FOV A as a front side event. The (45,30) and (135,30) CME become partial halo when latitude is changed to $30^\circ$.

\begin{figure}[ht!]
	\centering
	\includegraphics[width=5.5in]{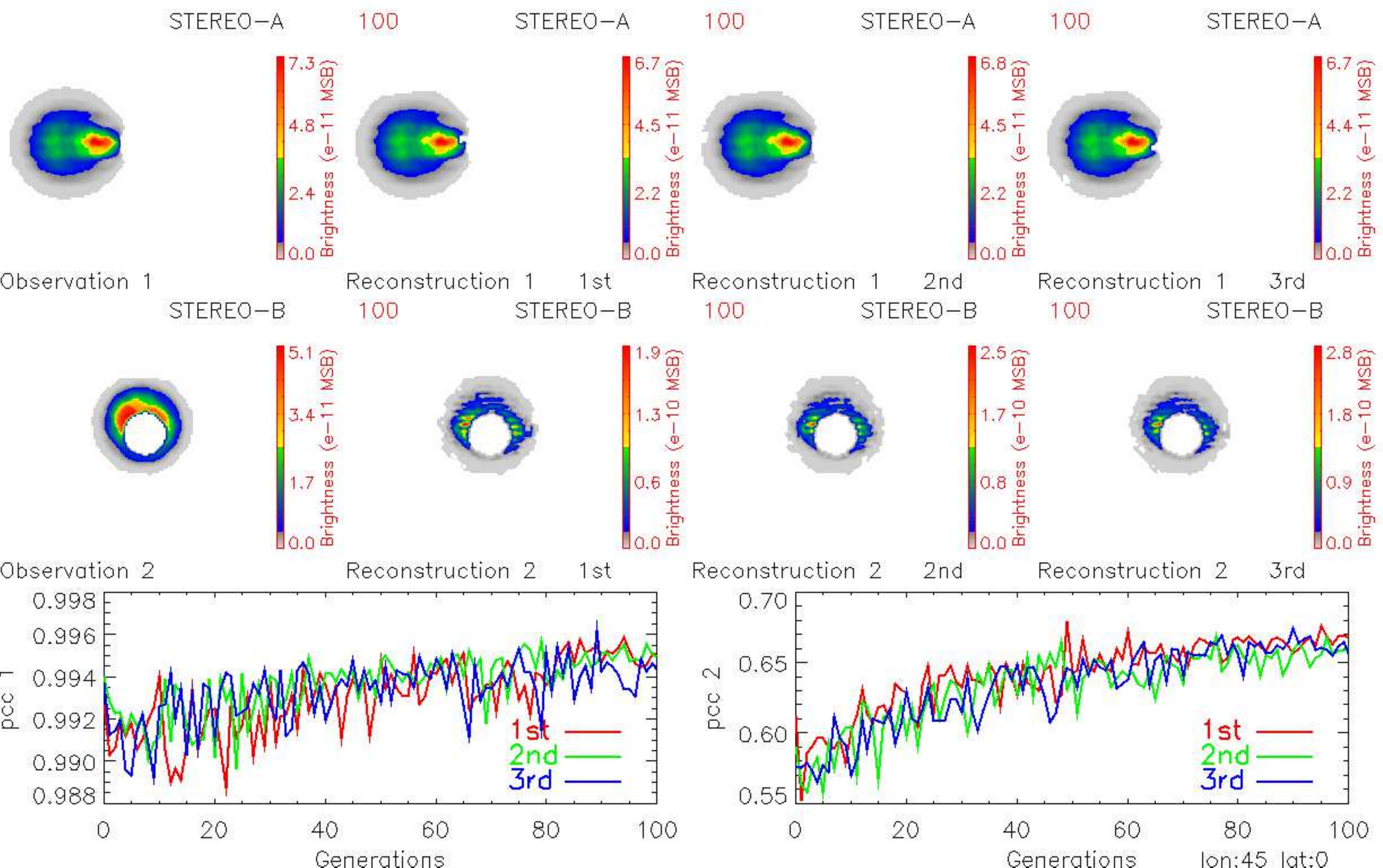}
	\\*
	\includegraphics[width=5.5in]{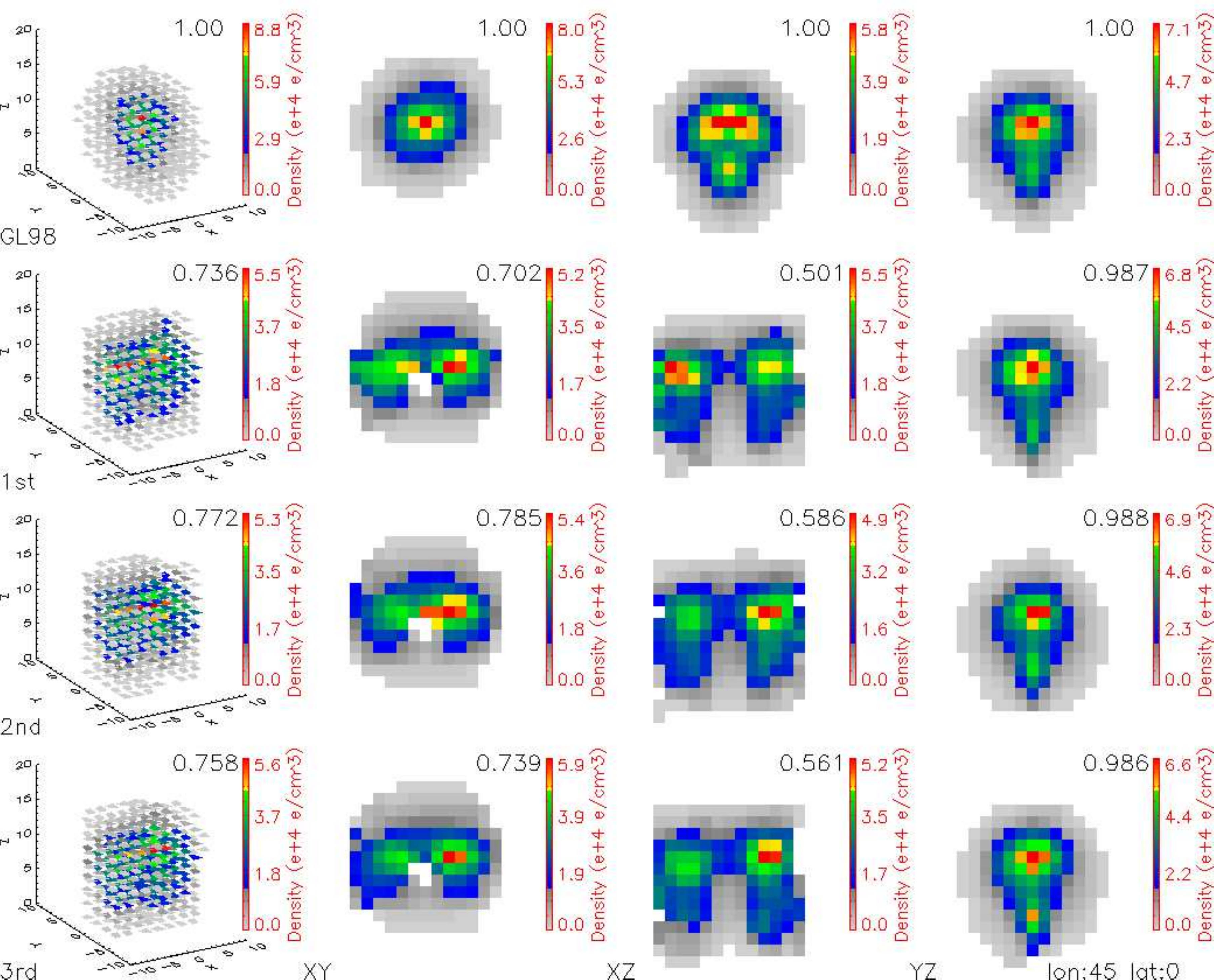}
	\caption{FOS of (45,0) CME with occulter. \hyperref[table_PCC]{Return to Table \ref{table_PCC}.}}\label{Fig_45_0_cut}
\end{figure}

\begin{figure}[ht!]
	\centering
	\includegraphics[width=5.5in]{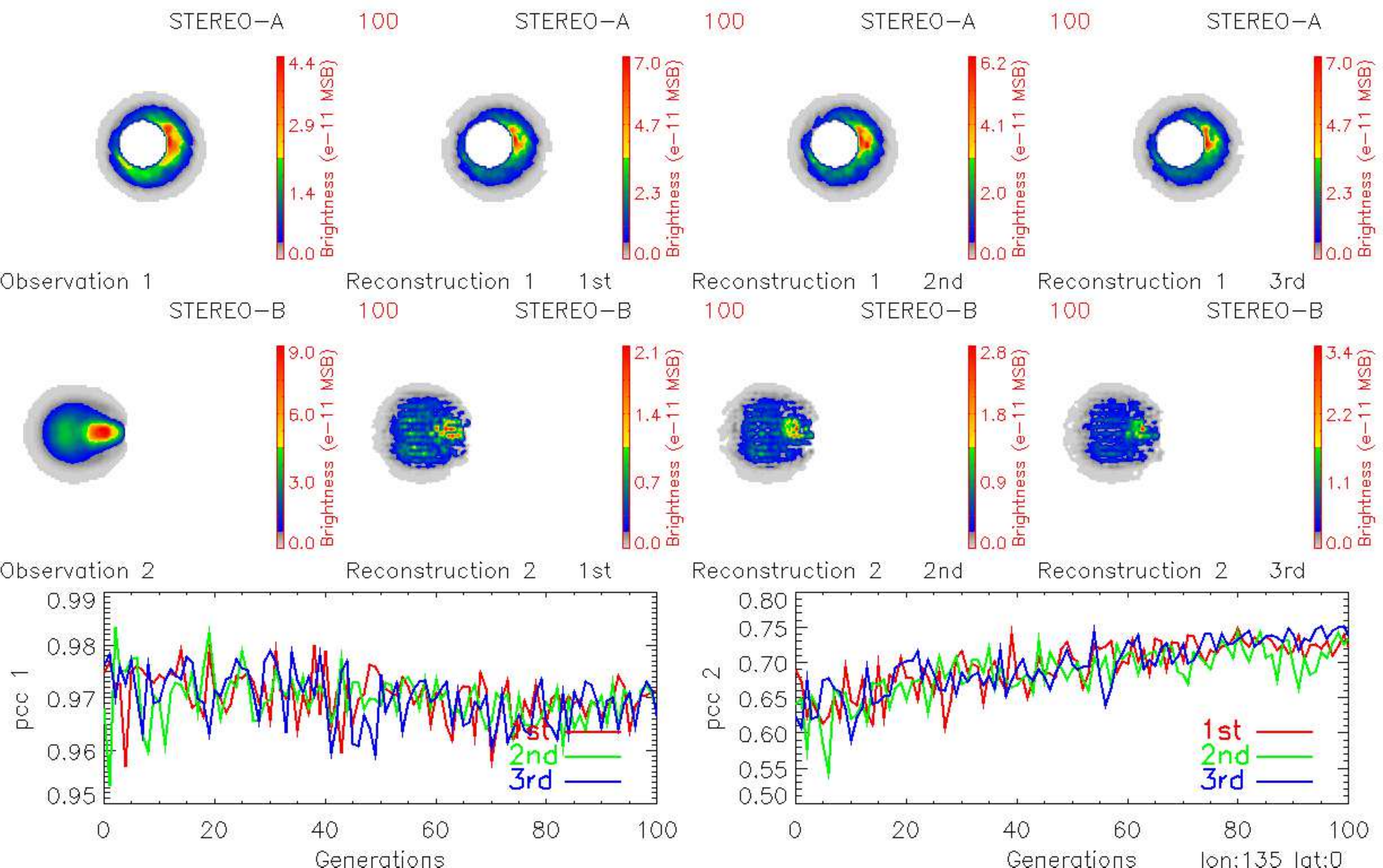}
	\\*
	\includegraphics[width=5.5in]{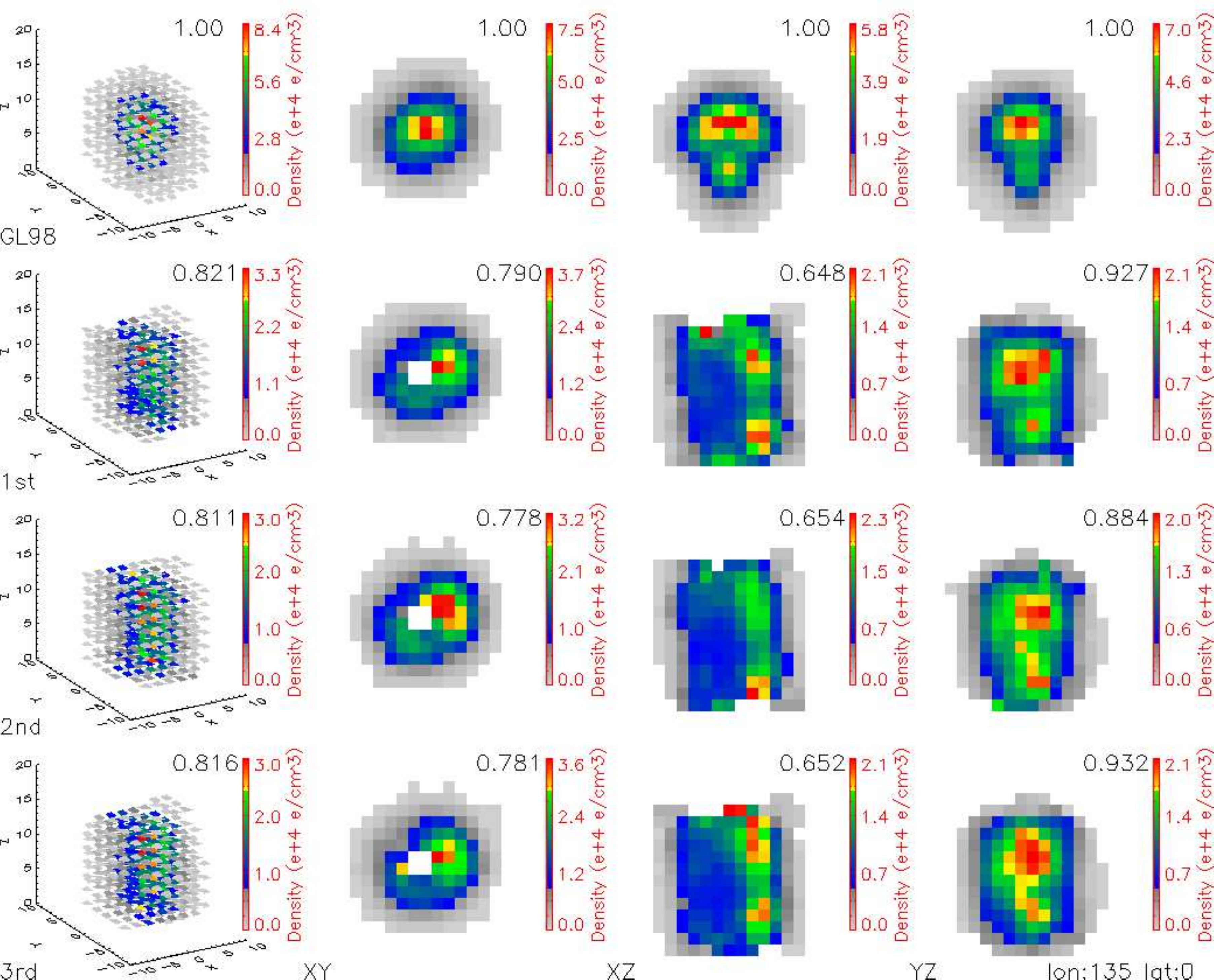}
	\caption{FOS of (135,0) CME with occulter. \hyperref[table_PCC]{Return to Table \ref{table_PCC}.}}\label{Fig_135_0_cut}
\end{figure}

\begin{figure}[ht!]
	\centering
	\includegraphics[width=5.5in]{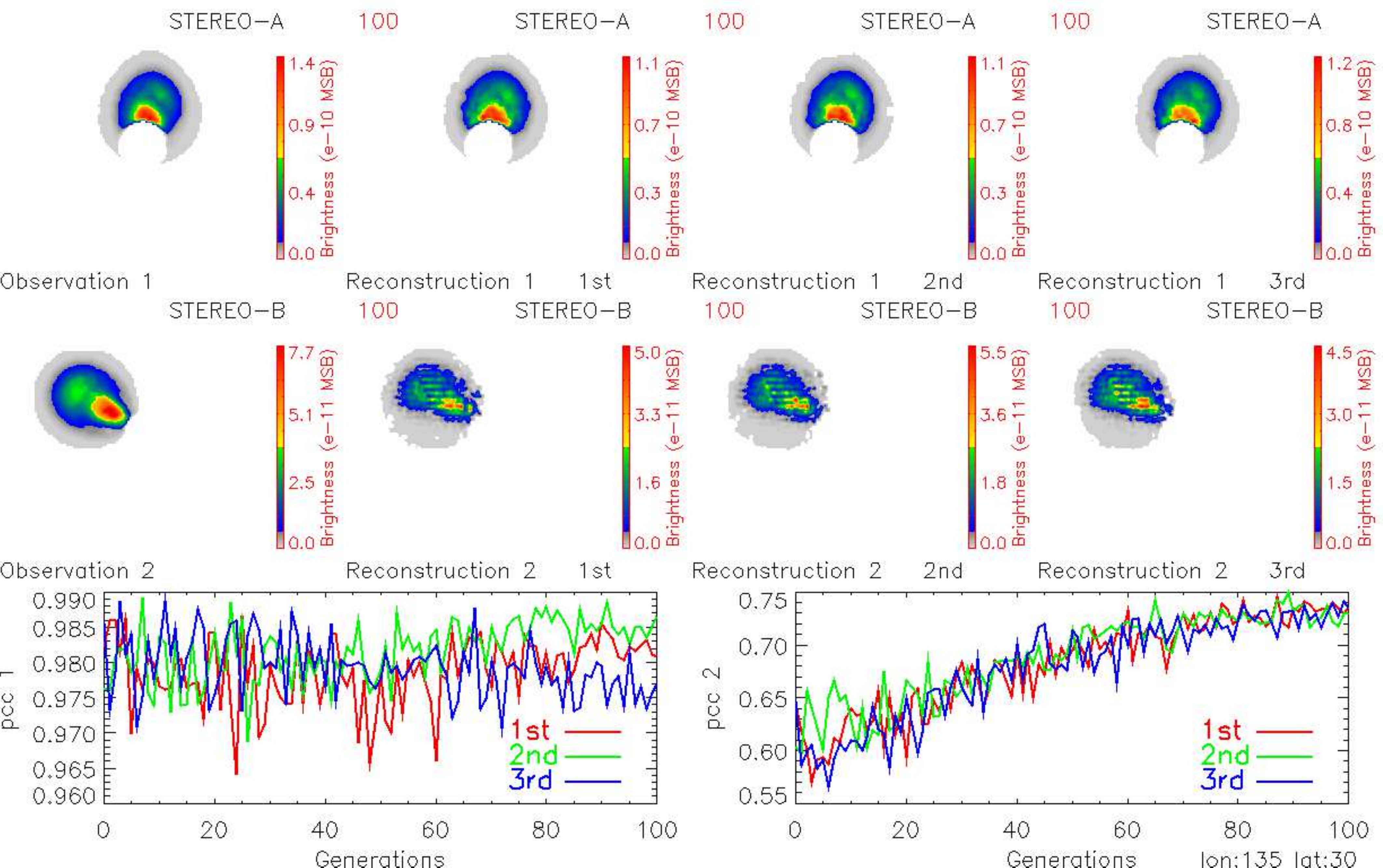}
	\\*
	\includegraphics[width=5.5in]{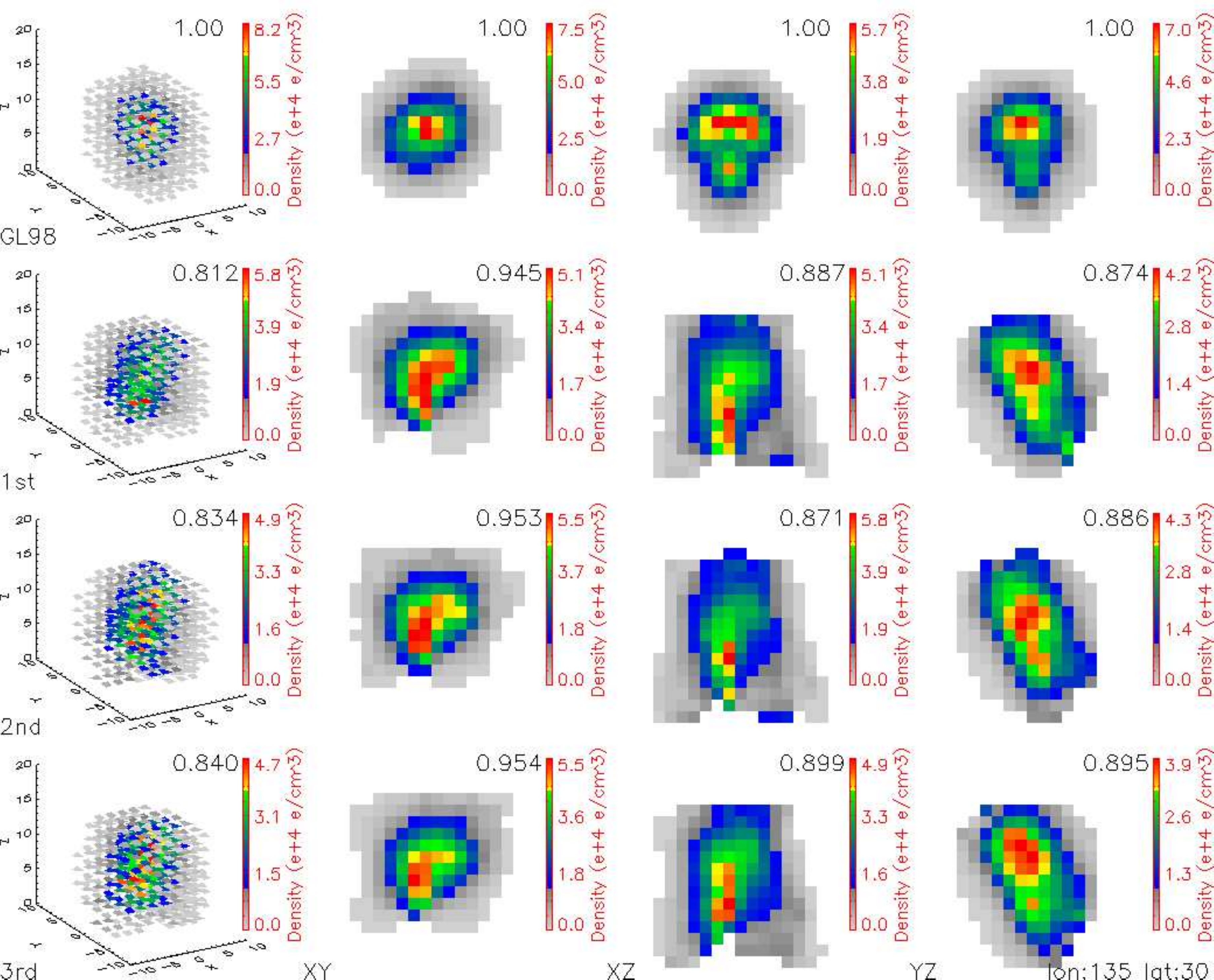}
	\caption{FOS of (135,30) CME with occulter. \hyperref[table_PCC]{Return to Table \ref{table_PCC}.}}\label{Fig_135_30_cut}
\end{figure}

The halo CMEs are obscured by occulter of the coronagraph as shown in Figure \ref{Fig_45_0_cut}, \ref{Fig_135_0_cut} and \ref{Fig_135_30_cut}. For halo CMEs, influence of occulter is more obvious than limb CMEs. Central part of the halo CME is hidden behind the occulter. For GRM reconstruction, situations in Figure \ref{Fig_45_0_cut} and \ref{Fig_135_0_cut} is different, although these two CMEs are both full halo. The (45,0) CME is full halo for observation 2 as shown in Figure \ref{Fig_45_0_cut}. On the other hand, the CME is almost not affected by the occulter since it is a limb event for observation 1. We obtain a relatively complete initial electron number of CME for GRM because the electron number is calculated from brightness in observation 1. In contrast, the initial electron number is obviously incomplete in Figure \ref{Fig_135_0_cut} for the (135,0) CME. 

As a result, brightness of reconstruction 2 is obviously larger than that of observation 2 for the (45,0) CME as shown in Figure \ref{Fig_45_0_cut} because of the completeness of CME in observation 1 and incompleteness in observation 2. Situation is opposite for the (135,0) CME as shown in Figure \ref{Fig_135_0_cut}.

\begin{figure}[ht!]
	\centering
	\includegraphics[width=5.5in]{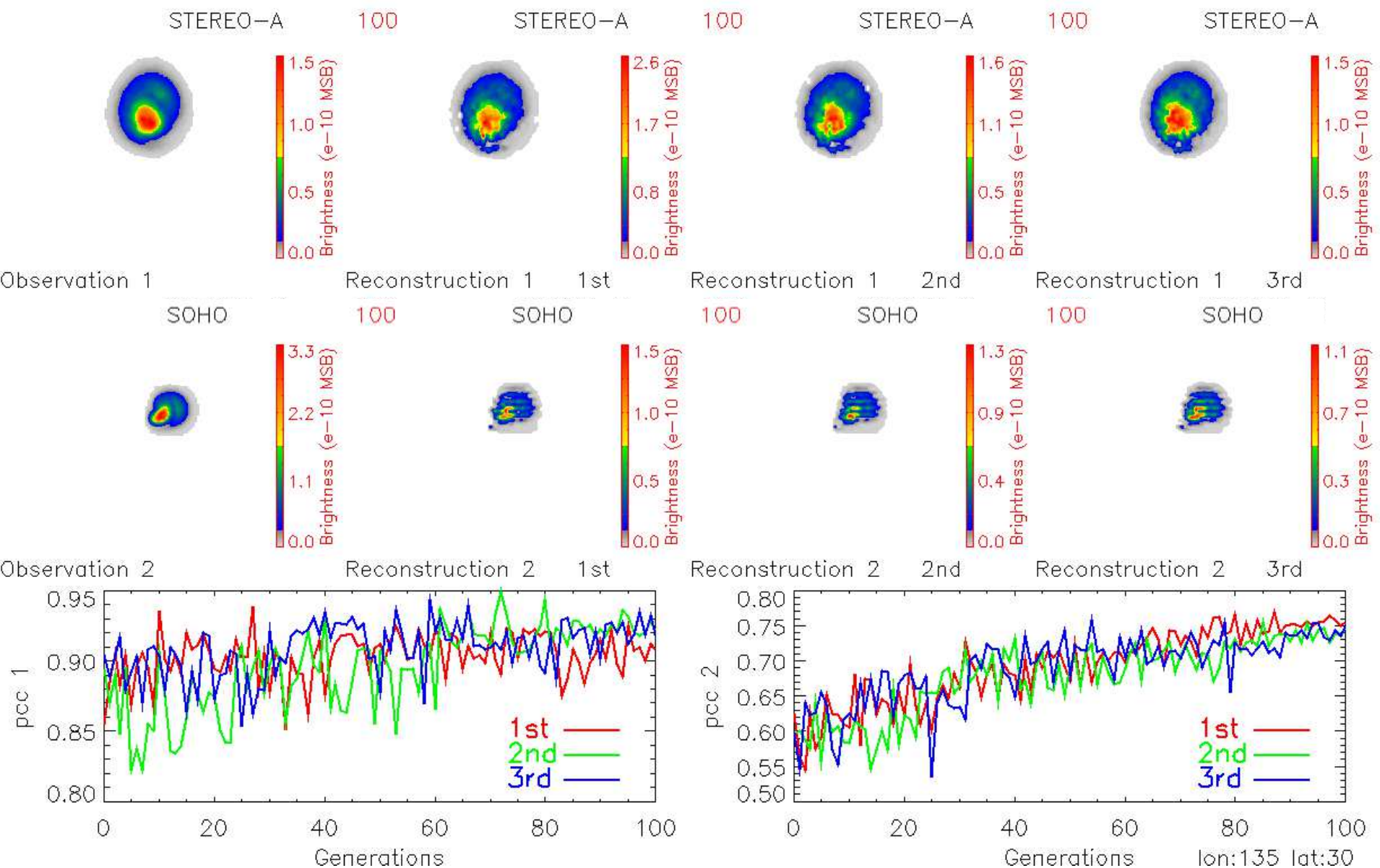}
	\\*
	\includegraphics[width=5.5in]{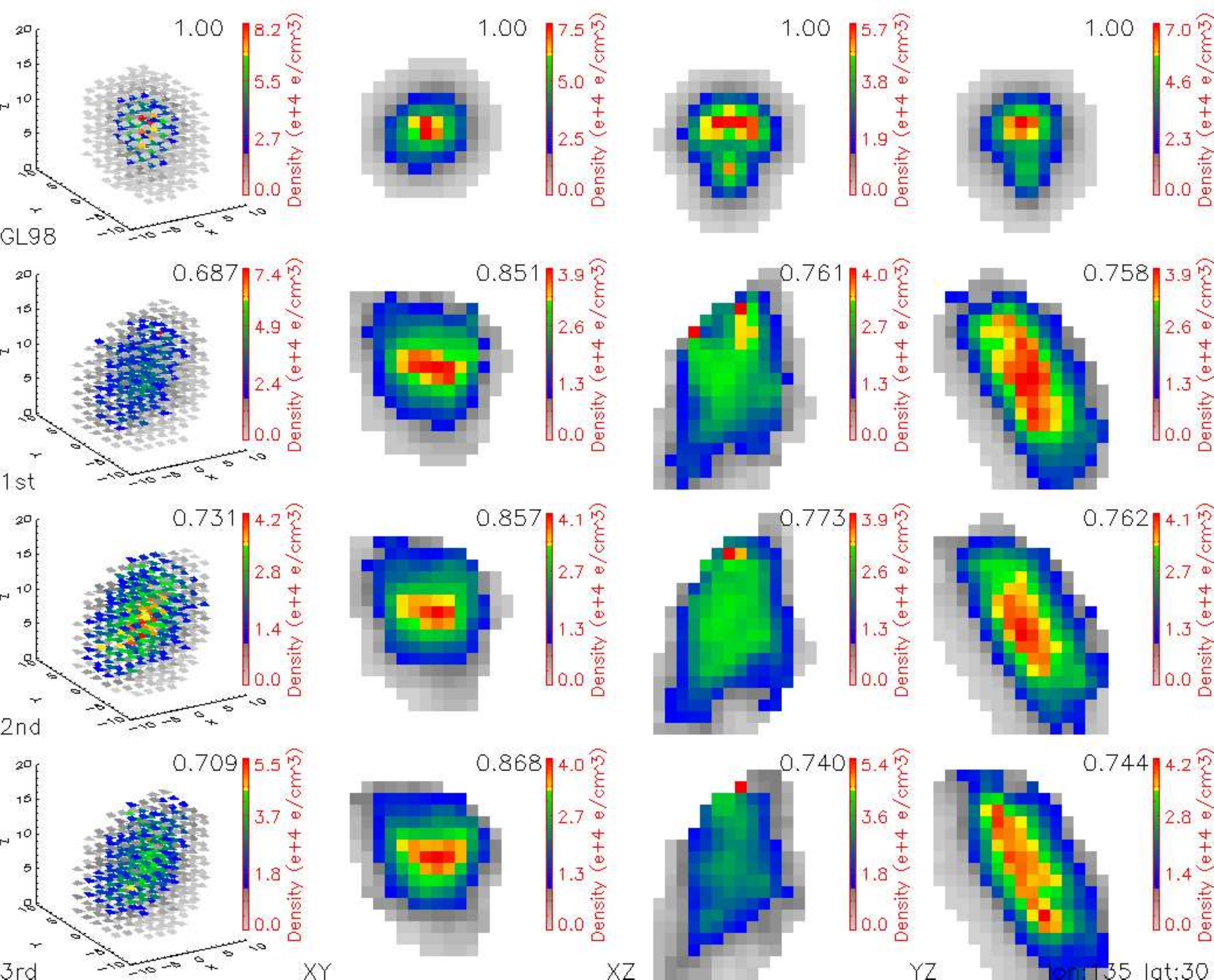}
	\caption{FOS of (135,30) CME using STEREO A and SOHO as observation 1 and 2. \hyperref[table_PCC]{Return to Table \ref{table_PCC}.}}\label{Fig_135_30_al}
\end{figure}

\begin{figure}[ht!]
	\centering
	\includegraphics[width=5.5in]{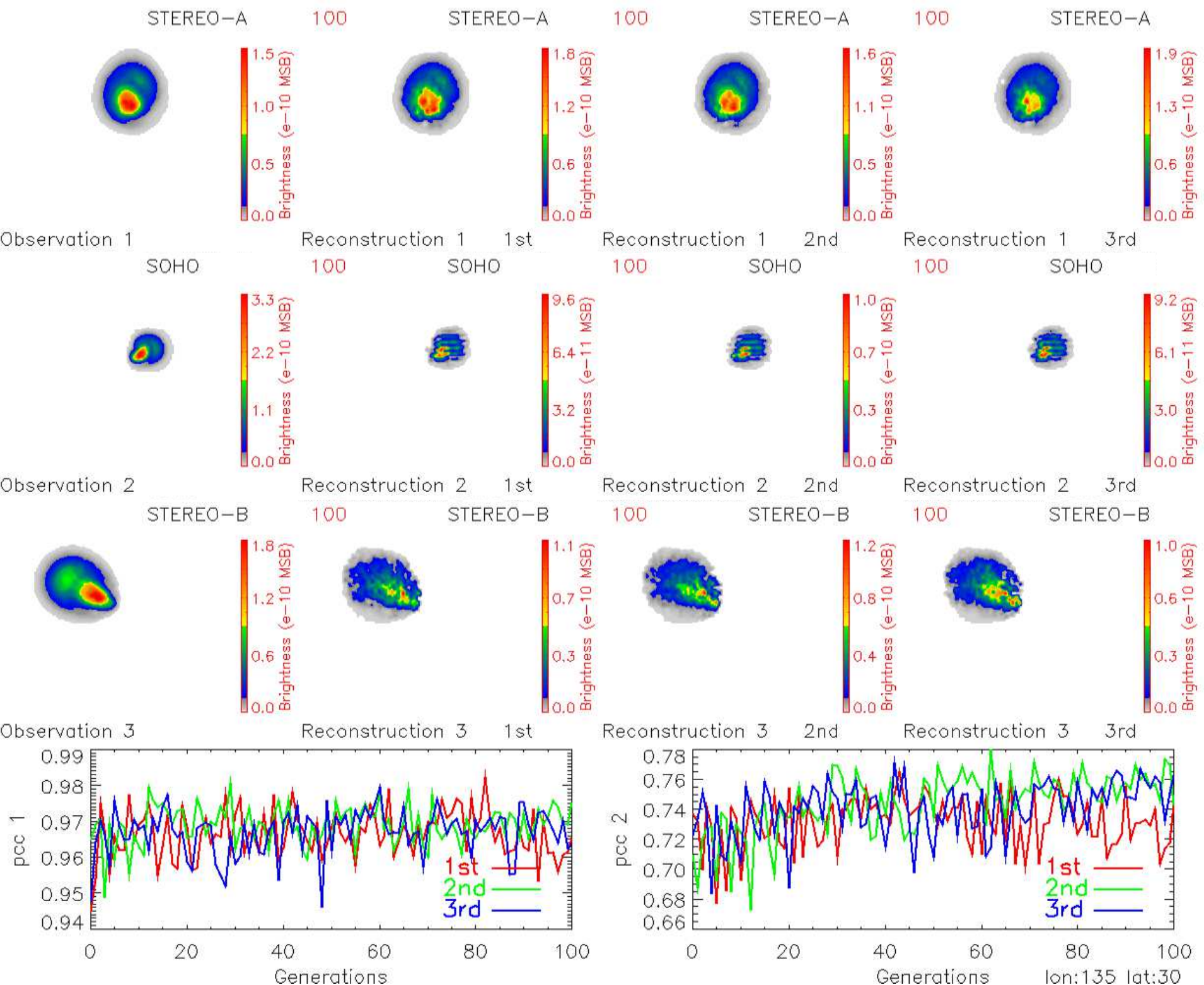}
	\\*
	\includegraphics[width=5.5in]{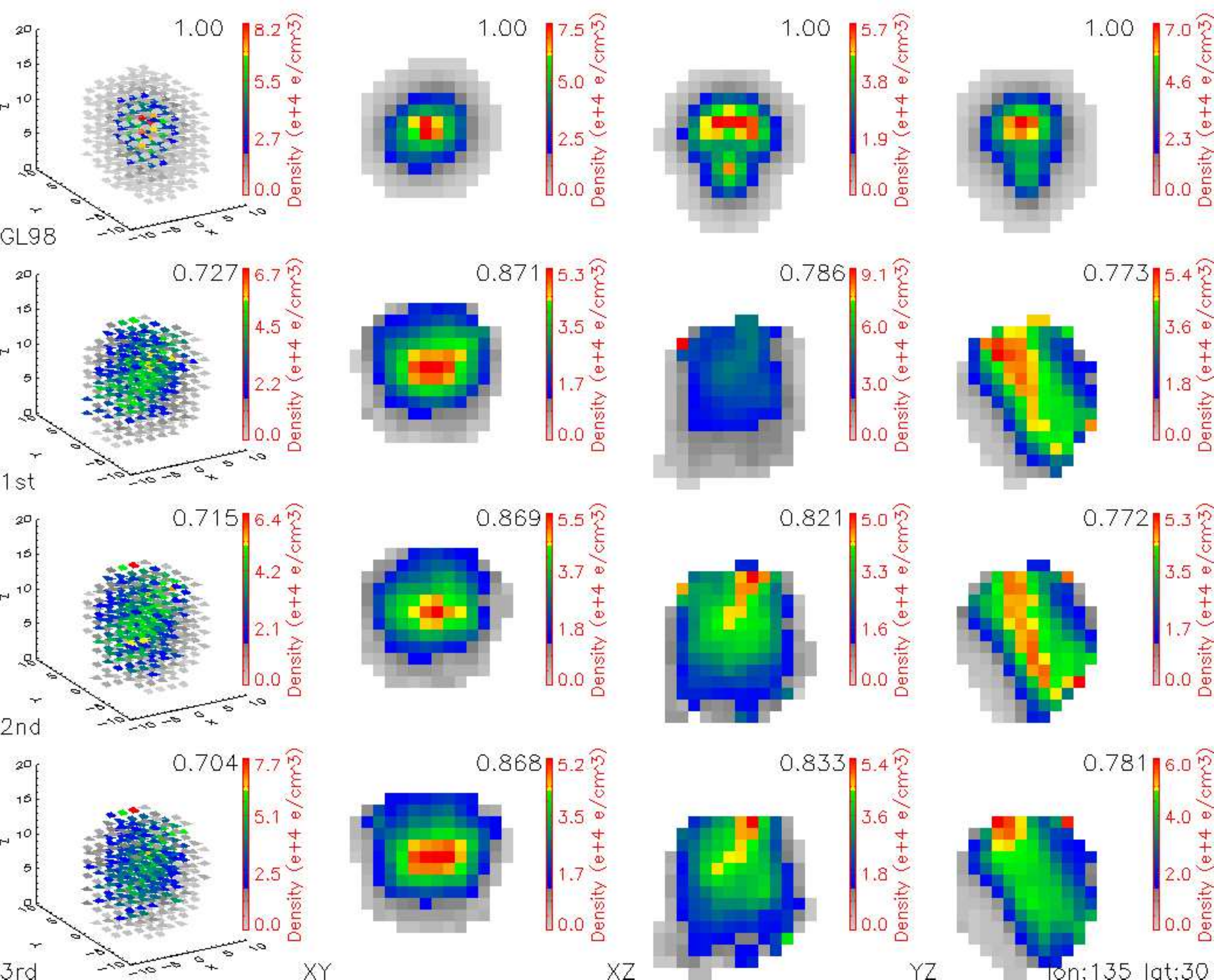}
	\caption{FOS of (135,30) CME using STEREO A, SOHO and B as observation 1, 2 and 3. \hyperref[table_PCC]{Return to Table \ref{table_PCC}.}}\label{Fig_135_30_alb}
\end{figure}

(3) As shown in Figure \ref{Fig_135_30_al}, the (135,30) CME is reconstructed using simulation data from view of SOHO/C3 as observation 2. Separation angle between STEREO A and SOHO is $135^\circ$. Comparing to the orthogonal coordinates with $90^\circ$ separation angle, $135^\circ$ separation angle is not ideal for GRM reconstruction. In order to improve the performance of GRM under non orthogonal coordinate system, view of STEREO B is added into the reconstruction as observation 3 in Figure \ref{Fig_135_30_alb}. Outline of the CME in observation 3 constrain the 3D electron distribution as shown in Figure \ref{Fig_135_30_alb} comparing to Figure \ref{Fig_135_30_al}. In the current work, we just use the outline of CME in observation 3 as a constraint. Such constraint makes the boundary of LOS more accurate. In the future, we may calculate the fitness function of  brightness in observation 3 together with those of observation 1 and 2 to optimize the reconstruction of electron distribution. 

\begin{figure}[ht!]
	\centering
	\includegraphics[width=3.5in]{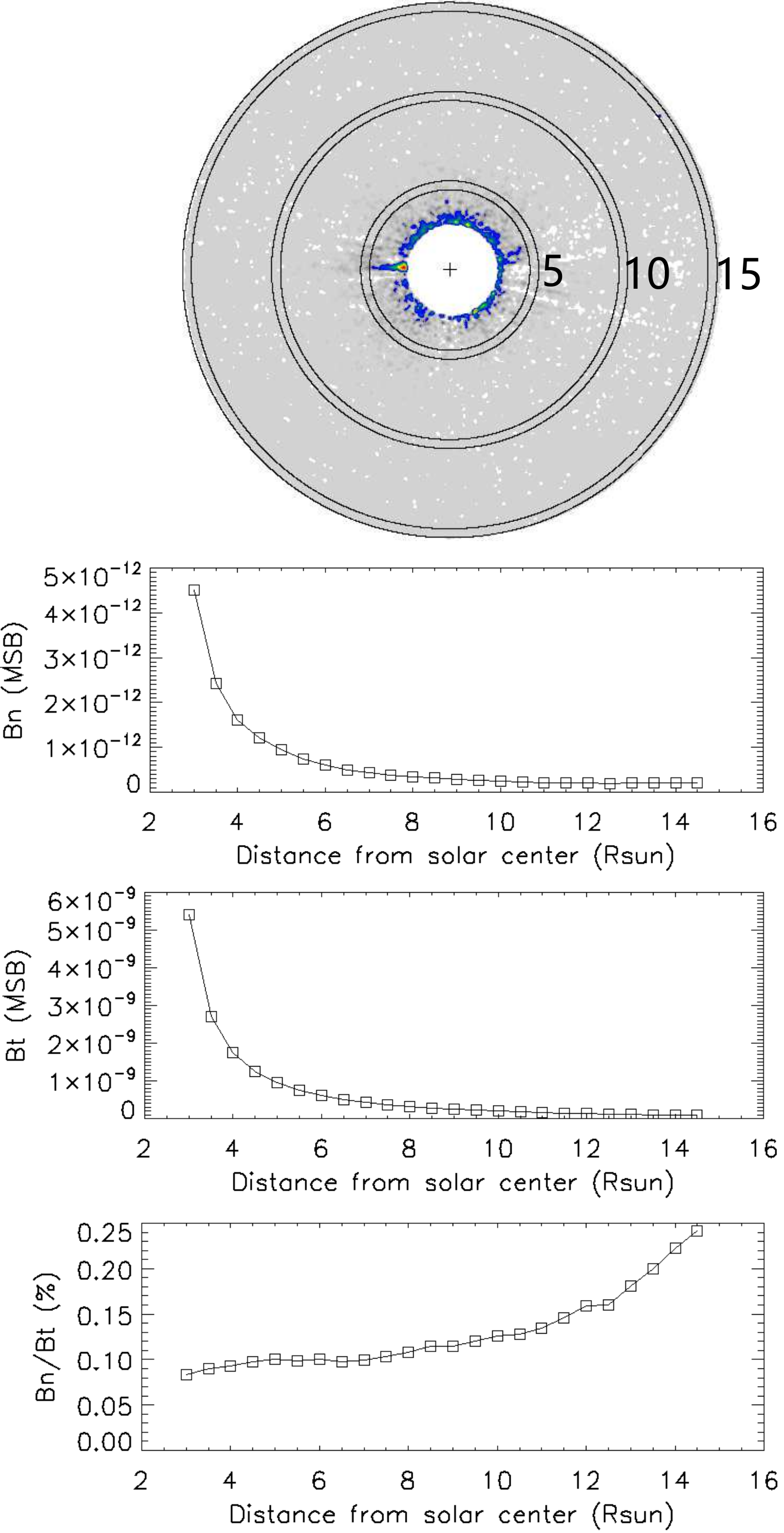}
	\caption{Noise level of COR2 image at 09:08 on July 10th, 2019 after subtraction of background image which is observed at 08:08. Top panel: Noise image. Centre of the Sun is marked with a plus. Rings with height of 5, 10 and 15 $R_\odot$ are labelled with number. Second panel: Plot of average noise in the rings, $B_n$. Third panel: Plot of average total brightness, $B_t$. Bottom panel: Plot of ratio of $B_n$ to $B_t$. \label{Fig_noise_plot}}
\end{figure}

\begin{figure}[ht!]
	\centering
	\includegraphics[width=5.5in]{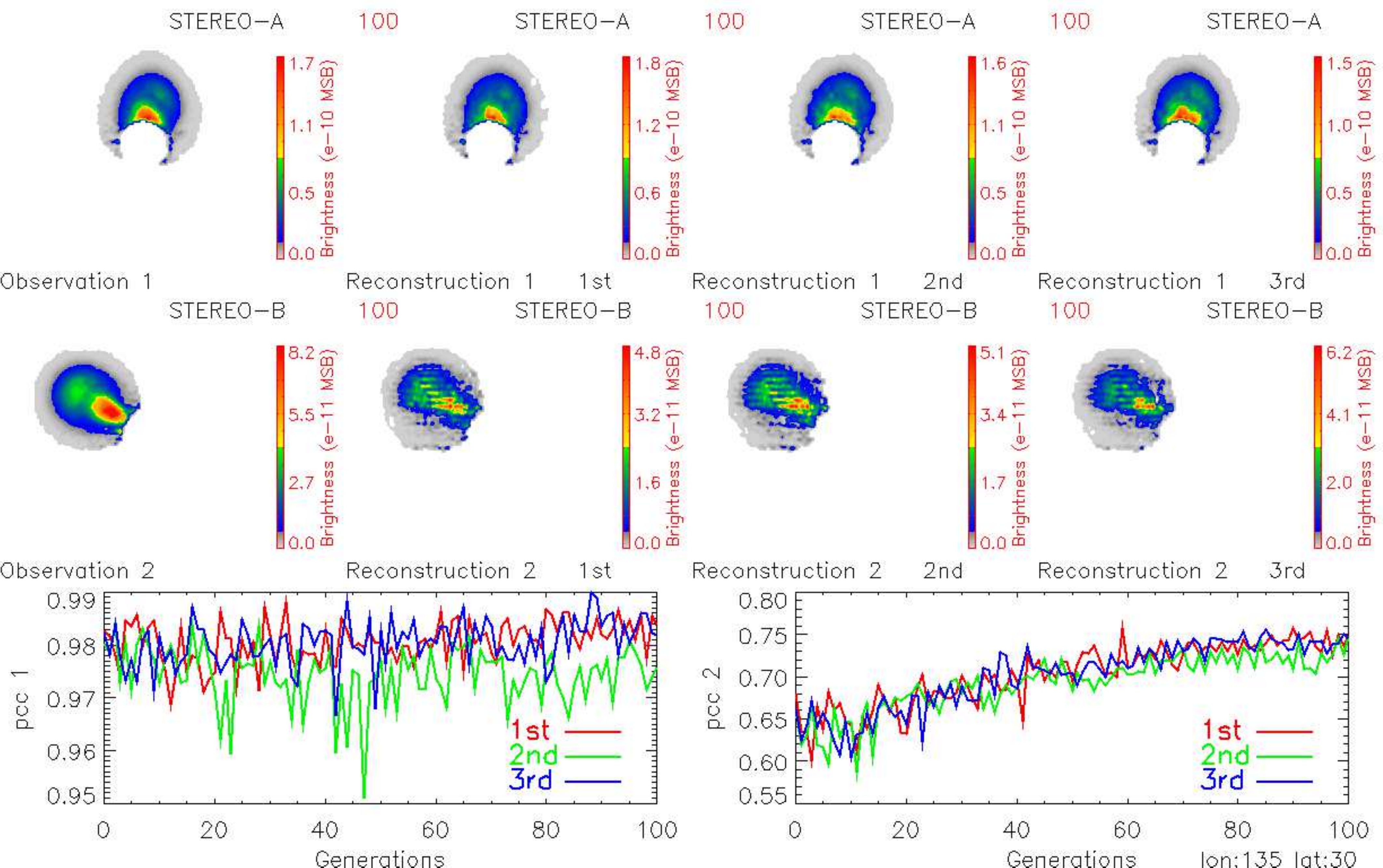}
	\\*
	\includegraphics[width=5.5in]{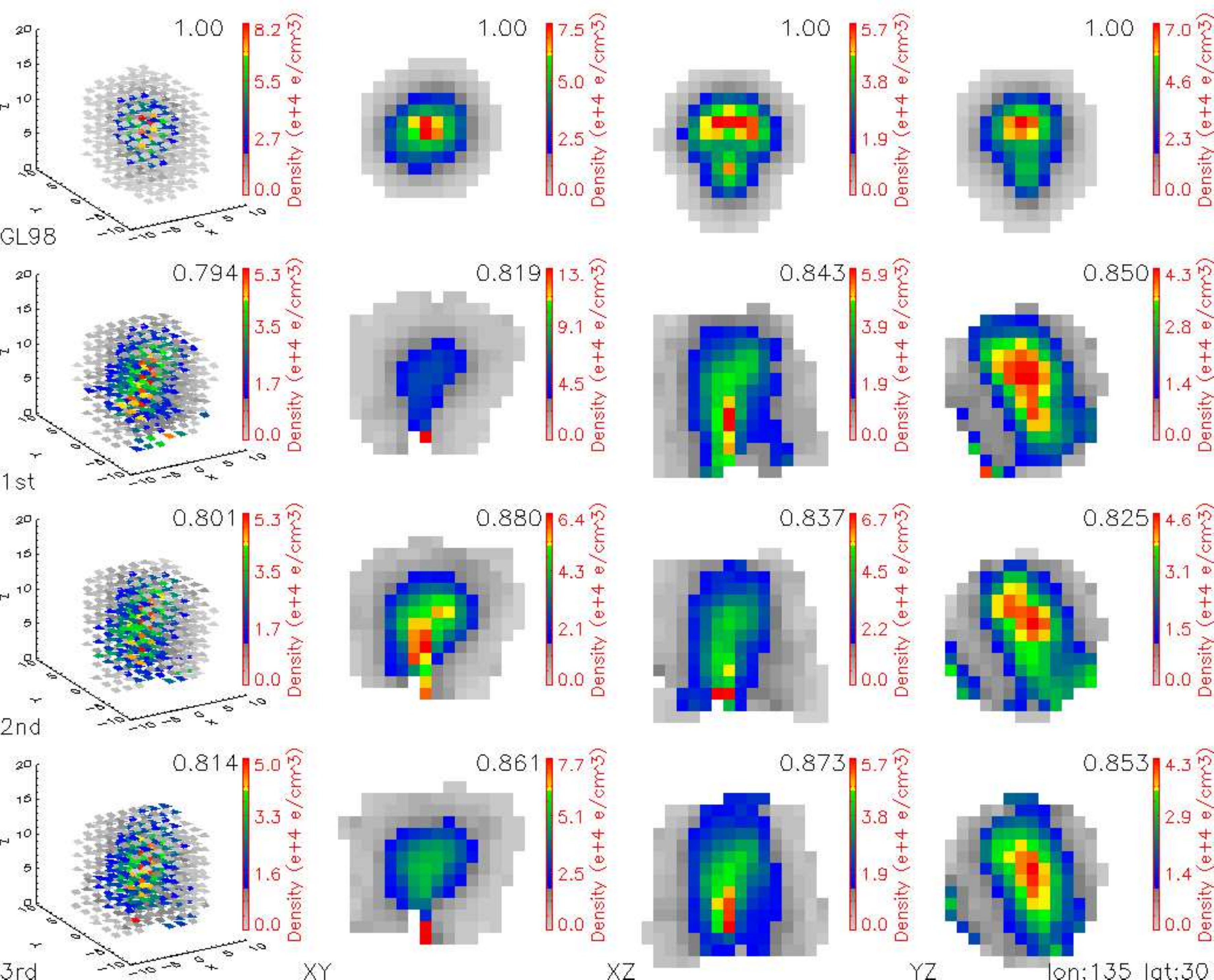}
	\caption{FOS of (135,30) CME with normal noise. \hyperref[table_PCC]{Return to Table \ref{table_PCC}.}}\label{Fig_135_30_noise_1}
\end{figure}

(4) For the original synthetic observations of model CME, there is no noise. By contrary, noise is inevitable in real observations. We add noise to the synthetic images of (135,30) CME as shown in Figure \ref{Fig_135_30_noise_1}. This is a critical test to evaluate the performance of GRM when the method is applied to real observations from coronagraphs. The noise image is produced by subtracting the background for a coronagraph image which is clear of CME. In a perfect subtraction of noiseless image, pixel values should be zero since there is no CME brightness. However, the brightness values usually deviate from zero for real observations as shown in the top panel of Figure \ref{Fig_noise_plot}. This deviation can approximate the normal noise level in real observations. In the second and third panel of Figure \ref{Fig_noise_plot}, average values of noise, $B_n$ and total brightness, $B_t$ are calculated within rings of 0.5 $R_\odot$ width. It is obvious that both of $B_n$ and $B_t$ decrease rapidly outward the Sun. However, the relative noise $B_n$/$B_t$ increase with height.

This normal noise level is added to the synthetic CME brightness as shown in Figure \ref{Fig_135_30_noise_1} and Table \ref{table_PCC}. Results of noise level multiplied by 10 and 100 are also shown in Table \ref{table_PCC}, respectively. We can see that difference of $PCC$ values between Figure \ref{Fig_135_30_noise_1} with normal noise and Figure \ref{Fig_135_30_cut} without noise is not obvious. In the case of normal noise level, values of $PCC$ are still larger than 0.8 and values of $MRAD$ are still less then 0.8. Difference becomes larger when the noise is multiplied by 10 and 100. An obvious decline of GRM performance can be seen in the case of noise multiplied by 100. In this case, values of $PCC$ are even less than 0.3 and values of $MRAD$ are close to 1.0. It means that the correlation between model and reconstructed CME is very low and GRM is no longer effective. On the other hand, GRM is still feasible under normal noise level which is common in the real CME observations.

\subsection{Application for observed CME}
\label{Observed_CME}
In this section, we show the application of GRM for a real CME event observed by STEREO and SOHO on September 30th, 2013. The CME is observed by COR2 on STEREO A and B at 00:08 and by C3 on SOHO at 00:04. Difference of the observation time is four minutes between COR2 and C3. The general case is that difference of observation timings usually exist between COR2 and C3. What we can do is to choose the nearest observation time of C3 to match the COR2 observation. Coordinate system of this GRM reconstruction is non orthogonal because separation angle between STEREO A and B is $73.5^\circ$ while separation angle between STEREO A and SOHO is $147.1^\circ$.

\begin{figure}[ht!]
	\centering
	\includegraphics[width=5.5in]{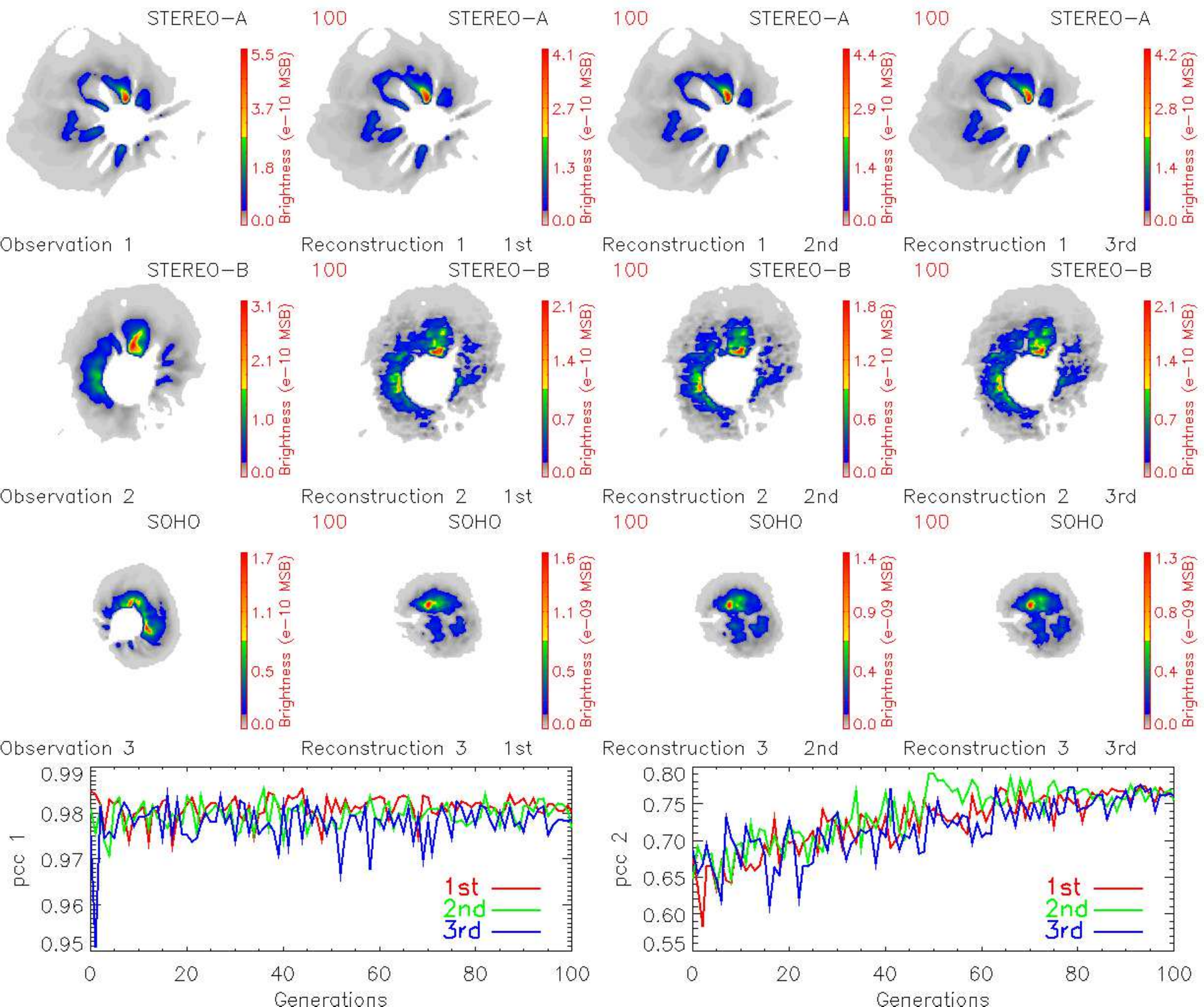}
	\\*
	\includegraphics[width=5.5in]{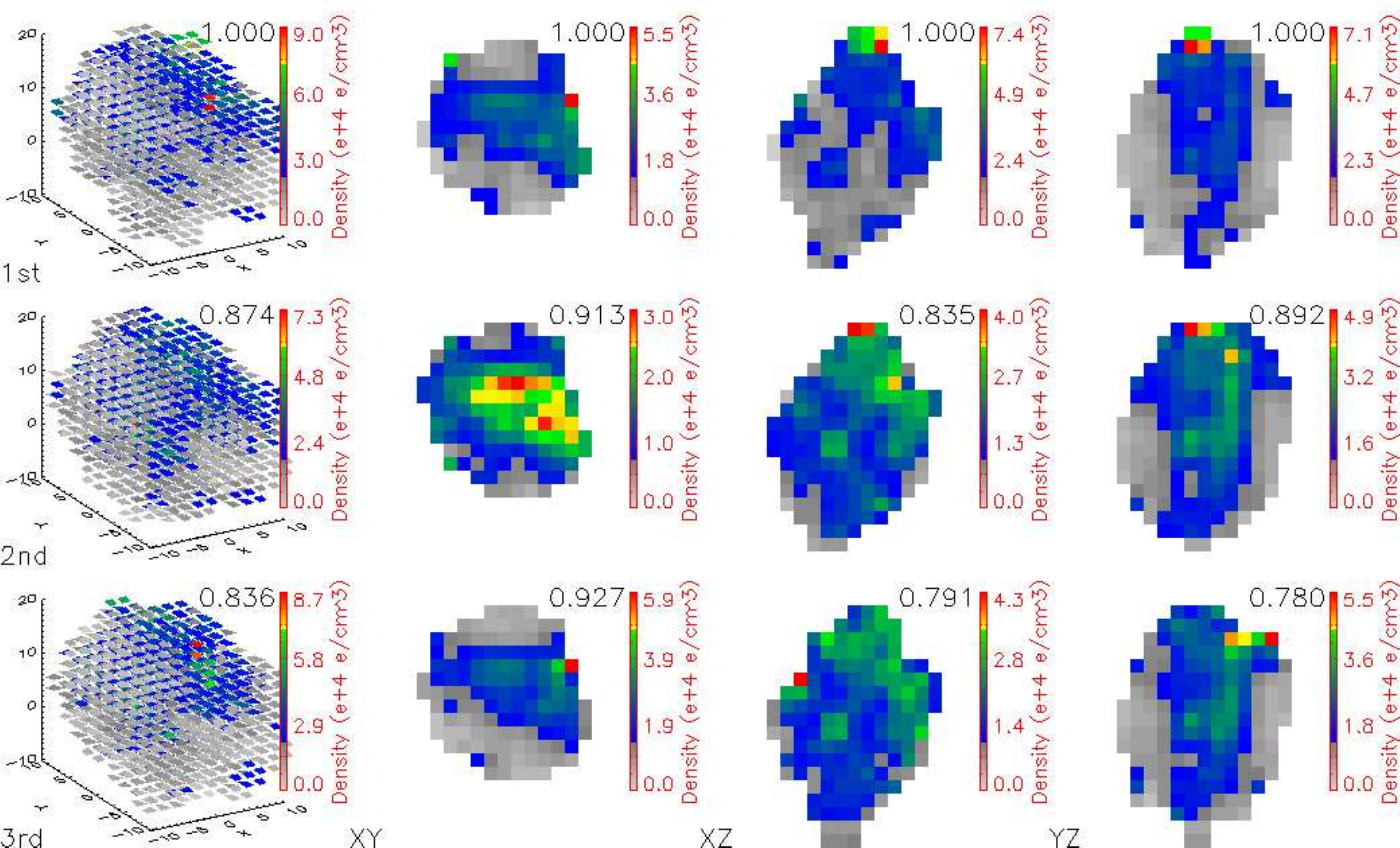}
	\caption{FOS of CME using real data from STEREO A, B and SOHO as observation 1, 2 and 3. \hyperref[table_PCC]{Return to Table \ref{table_PCC}.}}\label{Fig_abl_real}
\end{figure}

\begin{figure}[ht!]
	\centering
	\includegraphics[width=5.5in]{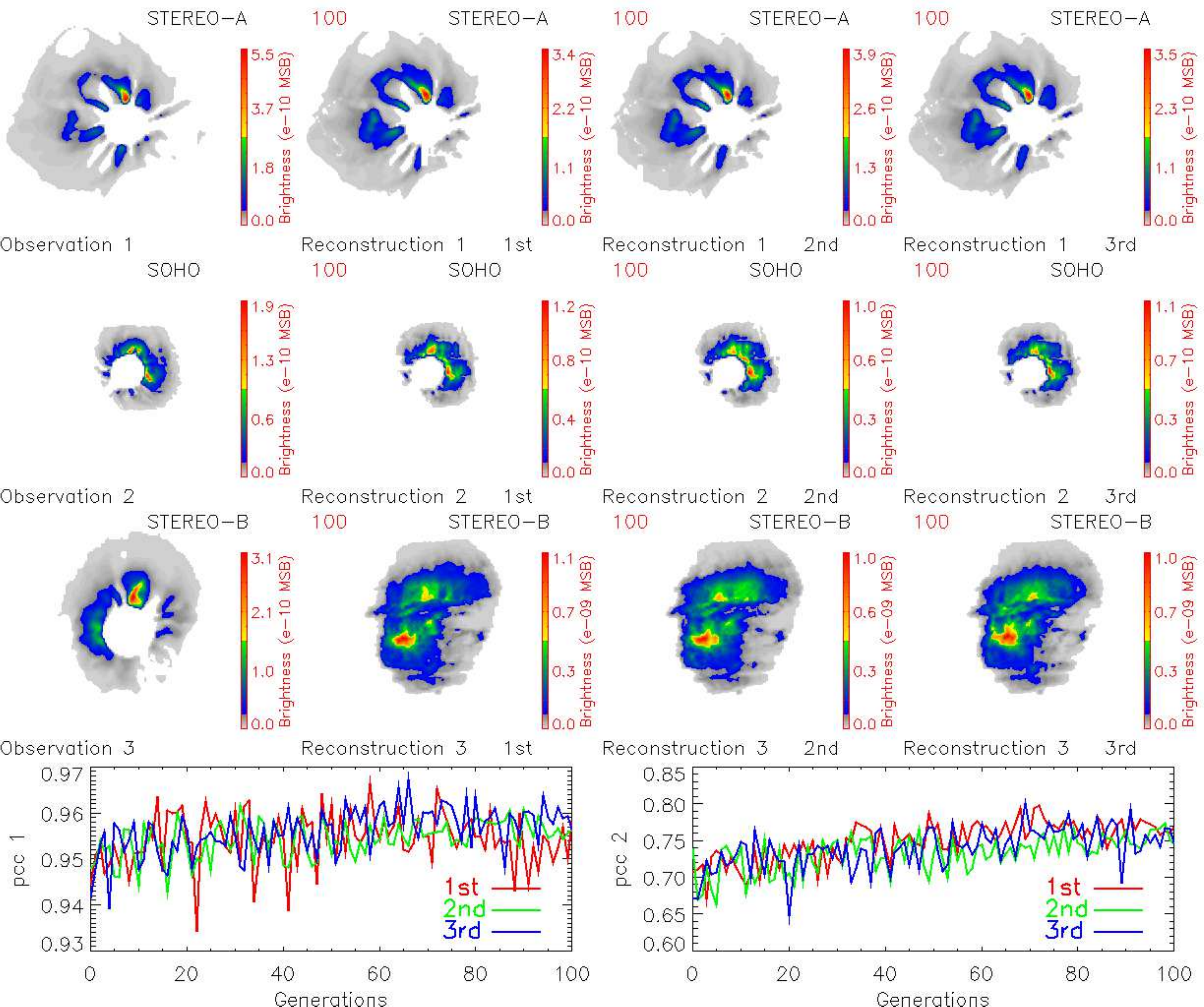}
	\\*
	\includegraphics[width=5.5in]{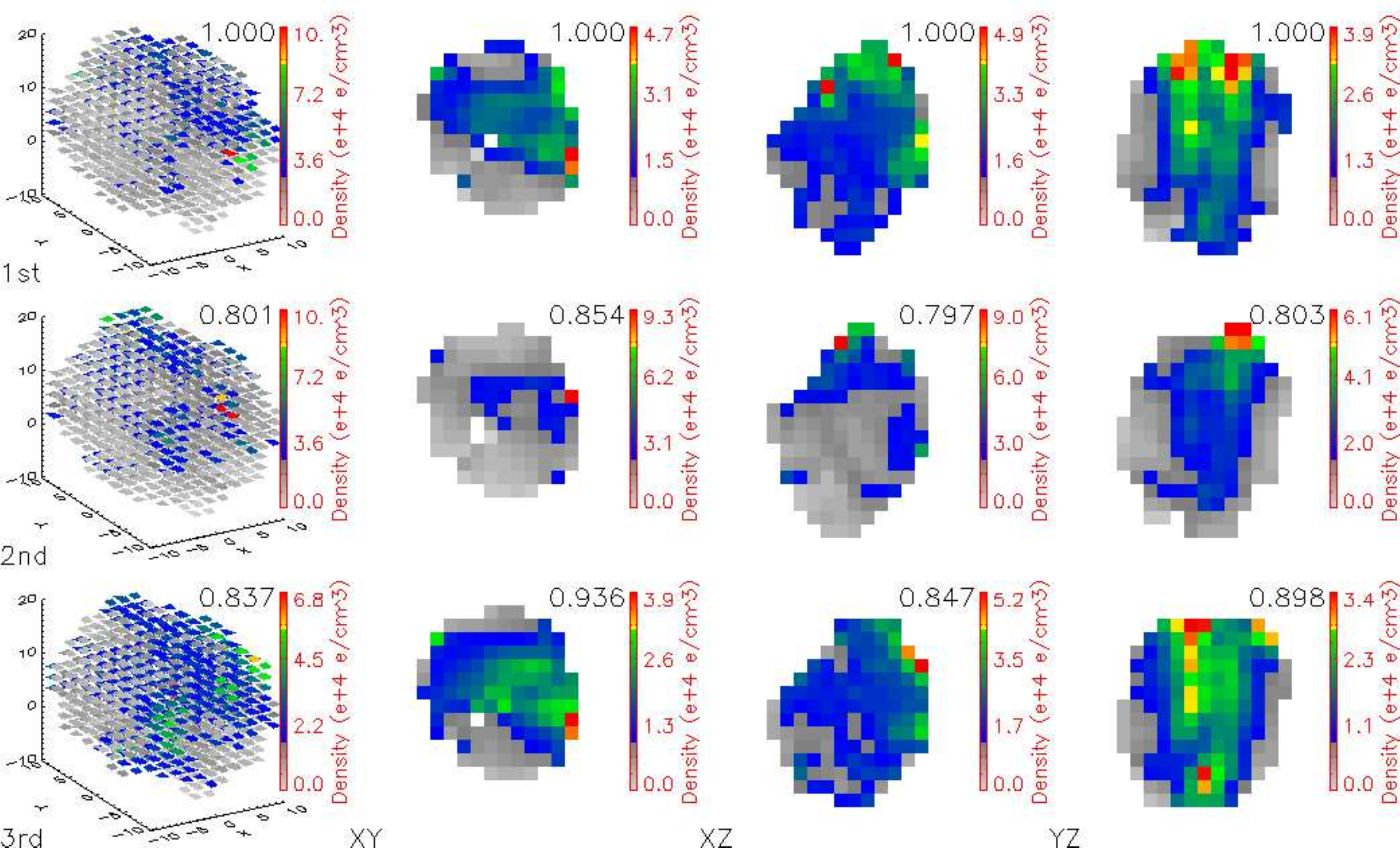}
	\caption{FOS of CME using real data from STEREO A, SOHO and B as observation 1, 2 and 3. \hyperref[table_PCC]{Return to Table \ref{table_PCC}.}}\label{Fig_alb_real}
\end{figure}

In Figure \ref{Fig_abl_real}, the CME is reconstructed from data of STEREO A, B and SOHO as observation 1, 2 and 3. In Figure \ref{Fig_alb_real}, we reconstruct the CME using STEREO A, SOHO and B as observation 1, 2 and 3. For this real CME, $PCC$ of brightness can still be calculated between real observation and reconstruction. Plots of the brightness $PCC$ from generation 0 to 100 still keep convergence as shown in the FOS. However, $PCC$ of electron distribution can not be calculated between real and reconstructed CME because we do not know the 3D electron distribution of the real CME. Instead, $PCC$ of electron distribution can be calculated for the ``2nd" and ``3rd" reconstruction by comparing to the ``1st" reconstruction. These values of $PCC$ are also summarized in Table \ref{table_PCC} marked with ``ABS" and ``ASB" for observation 2 using data from STEREO B and SOHO, respectively. Because we use density distribution in the ``1st" reconstruction as the reference (``model CME"), values of $PCC$ are equal to 1.0 and values of $MRAD$ are equal to 0.0 for the ``1st" reconstruction as shown in Table \ref{table_PCC}. For the ``2nd" and ``3rd" reconstruction, values of $PCC$ are larger than 0.78 and values of $MRAD$ are less than 0.4. These results show us the stability of reconstructions in different runs of GRM.

GRM reconstructs this real CME using data from STEREO B and SOHO as observation 2 respectively. The CME is partial halo in these two FOVs. Thus we can not detect the complete brightness of  CME core which is obscured by the occulter of coronagraph. On the other hand, the CME brightness is relatively complete in FOV A as observation 1 because the CME is a limb event in this FOV. A similar situation can be seen in Figure \ref{Fig_45_0_cut} for reconstruction of the model (45,0) CME with occulter. This model CME is a limb event in observation 1 and is a halo event in observation 2. Although the CME brightness is complete in observation 1, the reconstruction is still not correct because of the incomplete brightness in observation 2. The structure of reconstructed CME core is evidently destroyed as shown in Figure \ref{Fig_45_0_cut} comparing to the model CME. This may also be the reason that we can not find a typical dense core in the reconstruction of real CME as shown in Figure \ref{Fig_abl_real} and \ref{Fig_alb_real}. Besides the influence of occulter, the CME itself may not be a typical three part event. The dense core may become indistinct during the outward expansion.

As pointed out by \cite{2015ApJS..219...23M}, accuracy of reconstructed density depends on reliability of the input real data besides the reconstruction method. Calibration and background subtraction play important roles to obtain real CME brightness without contamination from vignetting, stray-light, F corona and other static structures in K corona like streamers and coronal holes, et al..

In this work, we download the level 0.5 fits files of coronagraph images and process them to level 1.0 using standard calibration routines from the Solarsoft library. For total brightness ($B_t$) observed by coronagraph C3 on board SOHO/LASCO, we apply the Solarsoft routine reduce\_level\_1.pro to make calibrations for dark current, flat field, stray light, distortion, vignetting, photometry, corrected time and position, et al.. For $B_t$ observed by coronagraph COR2 on board STEREO/SECCHI, we apply secchi\_prep.pro to make calibrations for subtracting the CCD bias, multiplying by the calibration factor and vignetting function, and dividing by the exposure time, et al.. 

After the standard calibration, $B_t$ contains a combination of K corona brightness ($B_k$) and F corona brightness ($B_f$). \cite{2015ApJS..219...23M} developed a method to separate $B_k$ and $B_f$ for C2 observations with FOV between 2.2 and 6.0 $R_\odot$. In this method, an in-flight calibration using star brightness is employed to improve the standard calibration for polarized brightness ($B_p$) of C2 which is usually obtained once a day. Series of calibrated $B_p$ during half period of solar rotation is subtracted from the corresponding $B_t$ to produce $B_f$ approximation. Finally, $B_k = B_t - B_f$. For C3 observations with FOV between 3.7 and 30 $R_\odot$, we may use this method to obtain cleaner corona brightness by removing $B_f$ from $B_t$ in the future.

$B_k$ contains brightness of Thomson scattering from coronal structure like streamers, coronal holes and CME, et al.. For CME reconstruction, we need to remove the background including the relatively static components like streamers, coronal holes. Subtraction of a pre-CME image observed before the CME is a standard approach as what we have done in this work. After such subtraction, the excess brightness of CME is available. \cite{2012ApJ...752..144M,2015ApJS..219...23M} developed a useful Dynamic Separation Technique (DST) to make the subtraction more effective. DST is valid under assumption that the quiescent structures like streamers are basically smooth in radial direction and slowly evolves in time while dynamic structures like CMEs are not radially smooth and evolve faster. Then DST is able to separate $B_k$ into quiescent and dynamic component using iterative deconvolution in the radial and time dimensions. In the future improvement, we will employ DST to subtract the background corona for CME reconstruction. \cite{2015ApJS..219...23M} presented a method to make cross calibration between C2 and COR2 which is necessary for electron density reconstruction using observations from different coronagraphs \citep{2019ApJS..242....3M}. Such method will also be considered in the future work.

\section{Conclusions}\label{sec:summary}
In this study, genetic reconstruction method (GRM) is used to reconstruct 3D distribution of CME electrons. We give a general description of CME reconstruction at first. Ill-posed property of the CME reconstruction is seriously pointed out. A set of analytic GL98 model CMEs with different orientations and shapes are employed to produce synthetic CME images without noise for the genetic reconstruction. Random noise is artificially added to some synthetic CME images to imitate the measurement noise which is unavoidable in the real CME observations. Since the electron distribution is known for the model CMEs, we make comparison based on $PCC$ for both of the electron distribution and its corresponding Thomson Scattering brightness between the model and reconstructed CME. $PCC$ of brightness from 0 to 100 generation presents the convergence of GRM for all of the model CMEs. $PCC$ of electron distribution in 3D space and 2D planes show us the ability of GRM to obtain stable and reasonable CME reconstruction. Values of $MRAD$ of electron distribution in 3D space validate the results of $PCC$.

Based on the comparison between model and reconstructed CME presented in FOS and Table \ref{table_PCC}, we understand more in depth about the advantages and limitations of GRM. Performance of GRM depends on the longitude and latitude of model CME as well as the completeness of observation and separation angle between points of view. A more reliable reconstruction can be obtained if: (1) The model CME in coronagraph is not much obscured by the occulter; (2) Central latitude of CME is about $30^\circ$; (3) Separation angle is about $90^\circ$. (4) Data of the third observation can be added into reconstruction.

Then the method is applied to real coronagraph data from STEREO A, B and SOHO. We compare the reconstructed brightness with observation to show the convergence of GRM. Comparisons of electron distribution between reconstructions from different runs tell us that the results of GRM reconstruction are stable. As pointed out in section \ref{Observed_CME}, we should employ more effective techniques to calibrate the coronagraph data and make the background subtraction in the future work.

The purpose of this paper is to demonstrate how GRM could be used to find a solution. Because of the ill-posed nature of CME reconstruction using data from only two or three view points and the random nature of GRM, results of GRM for the same model CME are different between three runs. This just illustrates the range of solutions which are possible for the unregularized problem. Regularization as in Equation \ref{eq_optimization} is helpful to further stabilize GRM reconstruction and may also mitigate the ill-posed characteristic towards a unique solution. How realistic the solution is then depends on how reasonable the regularization operator is. The GRM may allow least restrictions in the choice of the regularization. In the future work, we may test different forms of regularization to find suitable constraints for the CME reconstruction.

\begin{figure}[ht!]
	\centering
	\includegraphics[width=5.5in]{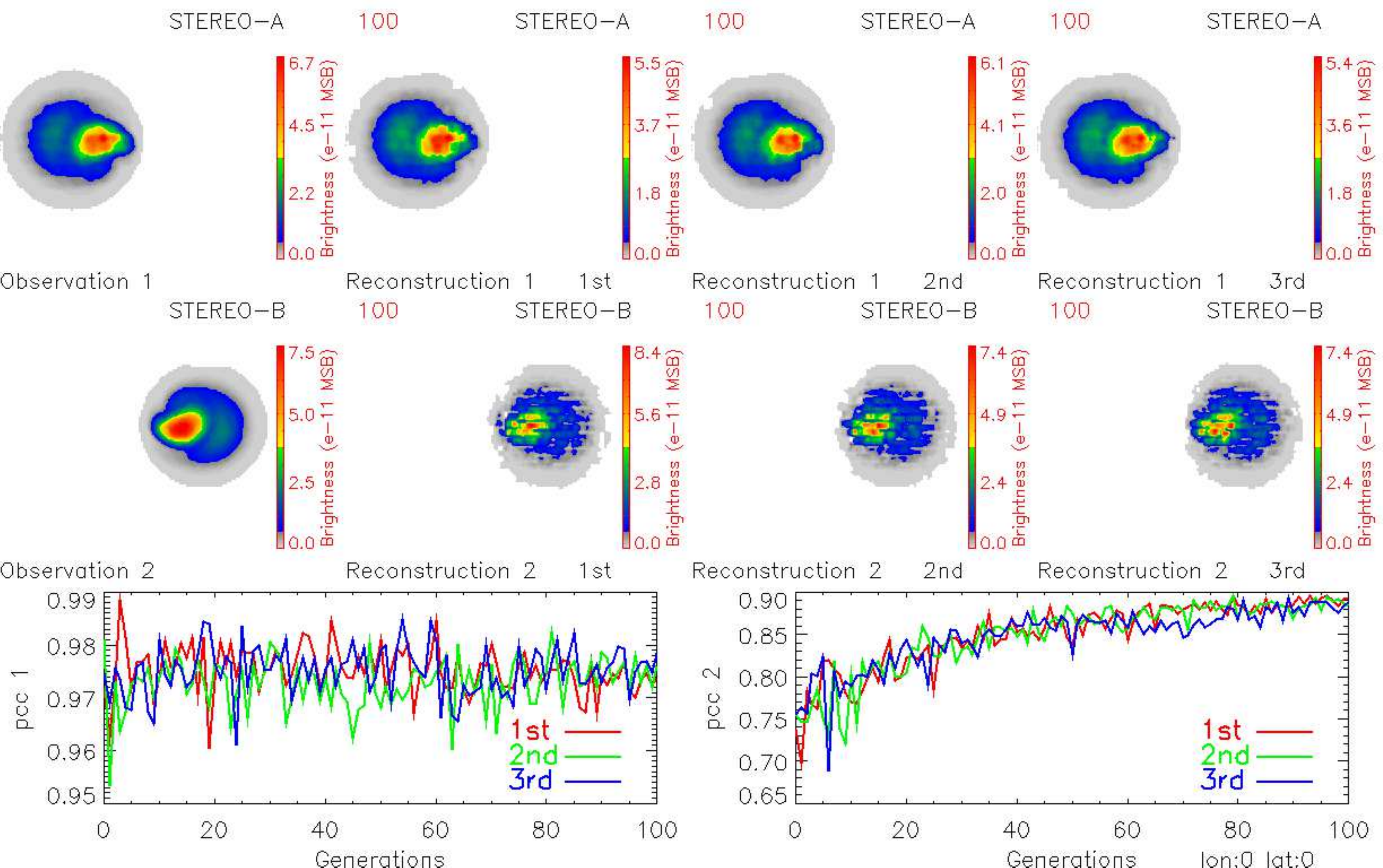}
	\\*
	\includegraphics[width=5.5in]{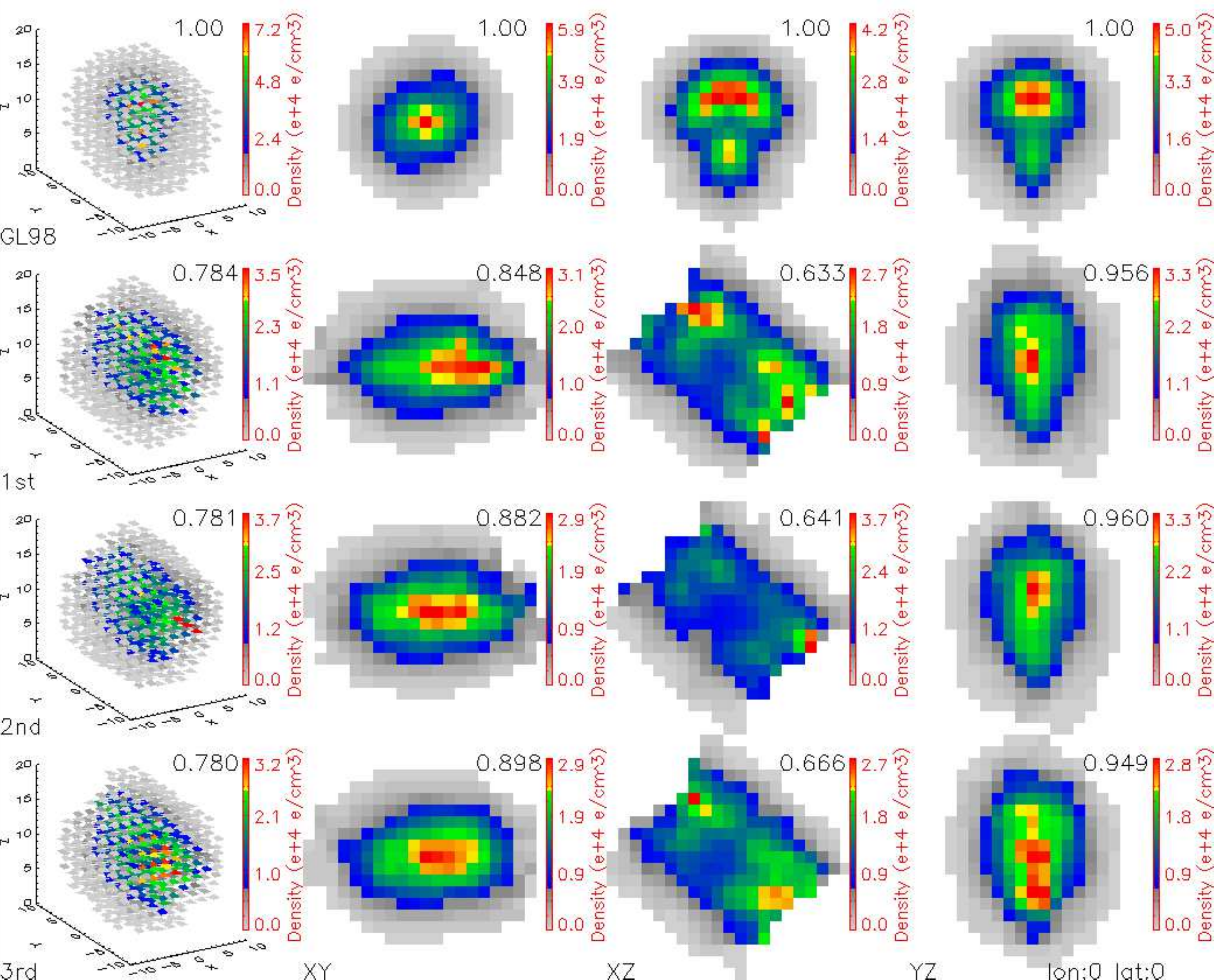}
	\caption{FOS of (0,0) CME. Top two rows: Brightness of model CME (observation 1 and 2) and brightness of reconstructed CME (reconstruction 1 and 2 of the ``1st", ``2nd" and ``3rd" GRM run) at generation 100. Third row: $PCC$ of brightness between model and reconstructed CME for observation 1 (left panel) and 2 (right panel) from generation 0 to 100 of the ``1st" (red), ``2nd" (green) and ``3rd" (blue) GRM run. Fourth row from left to right: Distribution of electron density of GL98 model CME in 3D space, xy plane, xz plane and yz plane. Bottom three rows: Distribution of electron density of reconstructed CME of the ``1st", ``2nd" and ``3rd" GRM run. The unit of electron density colour bar is $10^4$ $electron/cm^3$. \hyperref[table_PCC]{Return to Table \ref{table_PCC}.}}\label{Fig_0_0}
\end{figure}

\acknowledgments
We appreciate the anonymous reviewer for valuable comments which help us greatly improve the manuscript. We acknowledge SOHO and STEREO consortia for usage of the data. This work is jointly supported by the National Natural Science Foundation of China (NSFC) through grants 11603040, 11973058, 11473040, 11705210, 11873060 and U1831121.

\end{document}